\documentclass[sigconf]{acmart}
 
\usepackage{float}
\AtBeginDocument{%
  }
\usepackage{tabularx}
\usepackage{bbding}
\usepackage{graphicx}
\usepackage{geometry}
\usepackage{amsmath}
\usepackage{makecell}
\usepackage{float}
\usepackage{color}
\usepackage{booktabs}
\usepackage{caption}
\usepackage{tabu}
\usepackage{hyperref}
\usepackage{booktabs}
\usepackage{pifont}
\usepackage{subcaption}

\copyrightyear{2026}
\acmYear{2026}
\setcopyright{cc}
\setcctype{by}
\acmConference[DIS '26]{Designing Interactive Systems Conference}{June 13--17, 2026}{Singapore, Singapore}
\acmBooktitle{Designing Interactive Systems Conference (DIS '26), June 13--17, 2026, Singapore, Singapore}
\acmDOI{10.1145/3800645.3812941}
\acmISBN{979-8-4007-2563-0/2026/06} 

\author{Sijia Liu}
\authornote{Equal contribution.}
\email{sijialiu7-c@my.cityu.edu.hk}
\orcid{0000-0002-2502-0089}
\affiliation{
\institution{City University of Hong Kong\\Studio for Narrative Spaces}
\city{Hong Kong, SAR}
\country{China}}

\author{Hoi Ching Silvester Mok}
\authornotemark[1]
\email{hcmok7@cityu.edu.hk}
\orcid{0009-0005-4052-6490}
\affiliation{
\institution{City University of Hong Kong\\School of Creative Media}
\city{Hong Kong, SAR}
\country{China}}

\author{Long Ling}
\email{lucyling0224@gmail.com}
\orcid{0009-0001-2635-788X}
\affiliation{
\institution{Tongji University\\College of Design and Innovation}
\city{Shanghai}
\country{China}}

\author{Tobias Klein}
\email{ktobias@cityu.edu.hk}
\orcid{0000-0001-6426-0790}
\affiliation{
\institution{City University of Hong Kong\\School of Creative Media}
\city{Hong Kong, SAR}
\country{China}}

\author{RAY LC}
\authornote{Correspondences can be addressed to ray.lc@cityu.edu.hk.}
\email{ray.lc@cityu.edu.hk}
\orcid{0000-0001-7310-8790}
\affiliation{
\institution{City University of Hong Kong\\Studio for Narrative Spaces}
\city{Hong Kong, SAR}
\country{China}}

\begin{document}
\title[ClayScape]{ClayScape: A GenAI-Supported Workflow for Designing Chinese Style Ceramics with Clay 3D Printing}

\begin{teaserfigure}
 \centering
 \includegraphics[width=1\textwidth]{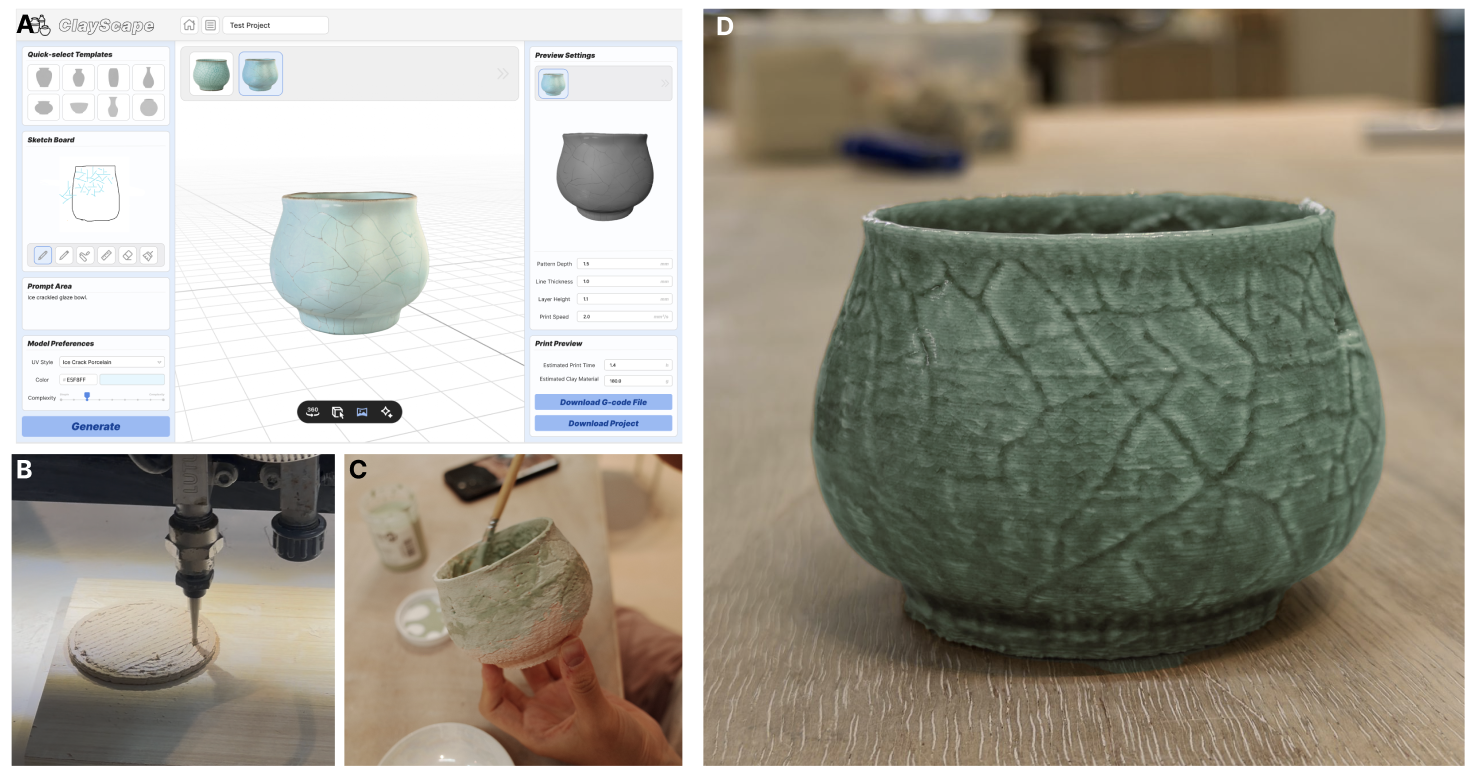}
 \caption{Hybrid fabrication workflow for creating a ceramic piece. (A) GenAI-driven tool ClayScape generates a 3D design from an initial sketch and text prompt. (B) Clay 3D-printing of the bisque piece. (C) Glazing process. (D) Final glazed ceramic work.}
 \label{introfigure}
\end{teaserfigure}


\begin{abstract}
Chinese ceramic-making involves complex and interdependent steps, making it technically demanding. Digital fabrication methods attempt to make the process more accessible, but for craft-creators, technical challenges such as CAD and CAM skills remain major obstacles. To address this, we designed a hybrid workflow that integrates Generative AI with clay 3D printing to support new creative possibilities. We evaluated the workflow through ClayScape, a design tool that operationalizes this approach, with four ceramic creators. Our findings show that the workflow supports accessible ceramic creation while revealing both expanded opportunities for creative exploration and challenges in balancing agency and control. This work demonstrates how hybrid workflows can lower barriers to digital fabrication while supporting creative possibilities in culturally grounded ceramic practices.

\end{abstract}




\maketitle

\section{Introduction}\label{sec:Introduction}
Chinese ceramics have a long and evolving tradition, with richly symbolic forms, textures, and glazes that span dynasties \cite{valenstein1989handbook}. In contemporary practice, many artists are actively exploring ways to incorporate traditional Chinese aesthetics into their creations, with modern making methods and technologies \cite{liu_craft_2024, jingexamining}. However, traditional ceramic making often involves a series of complex, interdependent steps, such as shaping, carving, painting, and glazing. The segmented nature of production makes certain styles and patterns difficult to reproduce or adopt by an individual, due the technical complexity and labor intensity of Chinese ceramic making \cite{jingexamining, Gowlland01072009}. Even if a craftsperson is skilled at one or several of the steps, it does not mean it is easy to complete the entire process. 

These challenges for craftsmen remain in many forms of handcrafts and art, including other ceramics traditions. Thus, prior work in HCI has examined how digital fabrication can reduce manual steps or introduce new workflows in ceramic making \cite{toka_adaptable_2023}, such as printing or slip casting textures instead of using traditional decorative skills \cite{fanni_PAVEL_2022}, or adapting 3D clay printing for culturally specific pottery practices \cite{silva_lovato_american_2025}. Computer-aided design (CAD) and computer-aided manufacturing (CAM) tools have also been gradually integrated into some Chinese ceramics creations \cite{xie2025using}. For Chinese ceramics, which often involve diverse forms and complex surface textures, these digital fabrication methods offer potential support for the creation process. However, adopting these technologies requires significant training, and a lack of prior knowledge in computational manufacturing presents another formidable barrier for craftspeople seeking to integrate digital fabrication into their practices.

Recent developments in Generative AI (GenAI) suggests its use as a potential tool for lowering the technical barriers of computational design by transforming conceptual idea and preliminary sketches into prototype visuals\cite{yang_ai_2022, han_when_2024}. Previous studies have explored applying GenAI to simplify traditionally complex steps in Chinese craft design processes \cite{wang_harmonycut_2025, tao_aifiligree_2025}. Beyond 2D design, GenAI models have also been applied in 3D design and modeling applications, enabling the generation of 3D models with UV textures \cite{lee_impact_2024, meshyai, tripo3d}. Together, these developments indicate that GenAI can support the design of culturally grounded patterns and forms in craft practices, suggesting its potential to help address the technical challenges ceramic creators face when working with CAD and CAM tools.

However, although digital fabrication can simplify complex manual craft processes and GenAI can assist in CAD/CAM design, there has been limited exploration of how these approaches can be integrated to support craft creation. In this context, we designed a hybrid fabrication workflow to support Chinese ceramic artists in both the design and implementation stages of creation. Our approach aims to lower the barriers to digital fabrication, increasing accessibility and efficiency, while preserving creators’ sense of agency and creativity within a hybrid craft process. The study was conducted in two phases. In the first phase, two researchers (one specializing in designing Generative AI workflow for creative support and the other in ceramic practice) used GenAI to generate ceramic designs from sketches and simple text prompts, along with corresponding 3D models and patterns for clay printing. These models were then printed with engraved surface patterns to guide glazing. In the second phase, we designed a tool ClayScape that integrates all stages of the first-phase process into a cohesive workflow. We invited additional craft practitioners to use this tool to create their own ceramic works, enabling us to evaluate the workflow’s usability and creative potential.

Our findings indicate that our designed hybrid workflow introduces new possibilities for ceramic practice while also revealing tensions.  Participants used the workflow to generate and produce Chinese-style ceramic forms with textured surfaces, highlighting its potential to support creative exploration in ceramic practice. At the same time, the workflow introduced practical and conceptual challenges, including material and feasibility limitations in clay printing, and balance between structured guidance and creative freedom. Beyond these technical aspects, the study also reveals how creators experience hybrid fabrication. While beginner crafts-people particularly benefited from accessible entry points for the creative process, experienced artists especially reflected critically on applications of complex digital fabrication skills in traditional crafts. These findings suggest that hybrid fabrication workflows should not only improve accessibility and efficiency, but also carefully balance technical support with creative agency, material engagement, and the cultural preservation of craft values.

The research contributes in the following ways:

\begin{itemize}
    \item \textbf{A hybrid fabrication workflow} that integrates Generative AI, clay 3D printing, and traditional glazing, lowering CAD/CAM barriers while supporting creative engagement in culturally grounded ceramic practices.
\end{itemize}

\begin{itemize}
    \item \textbf{ClayScape,} a design tool that operationalizes this workflow by combining AI-driven previews, textured 3D models, and clay print simulations to make hybrid fabrication more accessible.
\end{itemize}

\begin{itemize}
    \item \textbf{Empirical insights:} Two studies revealed both opportunities (e.g., efficiency, creativity, accessibility, and engagement) and challenges (e.g., material constraints, feasibility limitations, and creative trade-offs) in hybrid fabrication, demonstrated through the creation of four ceramic artifacts.
\end{itemize}

\begin{itemize}
    \item \textbf{Design implications:} We summary a set of insights for designing hybrid craft workflows, including the need to balance accessibility with creative agency, retain material engagement within digitally mediated processes, and integrate cultural knowledge without constraining creative expression.
\end{itemize}

\section{Background and Related Work}\label{sec:Relative Work}
\subsection{Chinese Ceramic Craft}
Chinese ceramics have a history of thousands of years and hold an important position in the history of world ceramics \cite{pope1958chinese}, recognized for their technical sophistication and artistic diversity \cite{valenstein1989handbook}. Their development produced a wide range of glazes and color palettes, along with intricate methods of production \cite{wood1999chinese}. One Chinese encyclopedia recorded that some types of porcelain required as many as 72 steps to complete, often involving a team of craftsmen working collaboratively \cite{song_tiangongkaiwu_taoyan}. Although not all ceramic works demand such extreme complexity, the core stages of production, such as throwing, molding, glazing, and firing, are crucial and require careful attention to guarantee successful outcomes. 
\cite{Gowlland01072009}. For an individual craftsperson to complete a ceramic piece independently, they must master all these foundational skills, which reflects the demanding nature of the craft.

The evolution of Chinese ceramics has been related to cultural and historical transformations, shaping them as both everyday utensils and cultural symbols \cite{Lin2021}. Each dynasty brought changes in technique, aesthetic preference, and material innovation \cite{Movement_pierson_2012}. Over time, the success rate of firing, the richness of glazes and colors, and the operability of handcraft methods all continued to advance. In recent years, computational methods and digital fabrication tools have been introduced into ceramic production in China, offering new possibilities to design, model, color, and replicate forms, as well as improving efficiency \cite{xie2025using}. However, many creators remain rooted in traditional handcraft practices. Their limited exposure to computational skills such as CAD and CAM means that digital fabrication has not yet become mainstream in Chinese ceramic making \cite{liu_craft_2024}.

\subsection{Ceramics in HCI Studies}
Recent years, HCI researchers have began to explore how computational method and digital fabrication technologies can be applied to ceramics, extending both traditional practices and craft education. One direction has focused on the re-materialization of traditional ancient ceramic artifacts through 3D scanning and AI \cite{elran_ghost_2022, elran_probabilistic_2024}. Another direction has developed virtual and mixed-reality environments to enrich ceramic experiences, including immersive Virtual Reality (VR) environment for Canton porcelain appreciation \cite{YI_Immersive_2024}, and AI-augmented mixed-reality system for wheel-throwing practice and craft learning \cite{ji_reshaping_2025}. 

There are also projects that have explored new methods for surface decoration, including using computational fabrication systems to extend manual ornamentation \cite{toka_adaptable_2023}, adapting slip casting into a digital approach to create ceramic objects with intricate textured surfaces \cite{han_slip_2025}, incorporating glazed ceramic ware with electronic circuits for interaction \cite{zheng_crafting_2023}, applying CAD together with laser machining to process crackle patterns \cite{dick_design_2018}, using CNC tools for decorative engraving \cite{zoran_Hybrid_2015},and applying innovative ceramic materials to create animated surface effects \cite{ye_viviclay_2025}, as well as adopting collaborative robotic systems to support skilled artisans in adding texture to clay surfaces \cite{WORM}.

Hybrid fabrication approaches have increasingly involved 3D printing techniques \cite{tao_4doodle_2023, Takahashi_3d_2019, DigitalJoinery, PlayingthePrint}, which has contributed to the growing interest in clay-based digital methods \cite{toka_practice-driven_2024}. Clay 3D printing has gradually become an active area of research, with projects exploring both tools and materials. Related previous projects include the Digital Pottery Wheel, a hybrid wheel with 3D printing capabilities \cite{Throwing_Moyer_2024}; CoilCAM, a CAM programming system that generates parametric forms and textures \cite{bourgault_coilcam_2023}; WeaverSlicer, a slicer optimized for clay printing \cite{friedman-gerlicz_weaveslicer_2024}; and SketchPath, which allows artists to design hand-drawn toolpaths for clay printing \cite{frost_sketchpath_2024}. Other related studies have investigated new material properties for shape-changing clay print composites \cite{bell_shape-changing_2024}, examined the use of clay printing in culturally specific contexts such as American Indian pottery practices \cite{silva_lovato_american_2025}, and explored emerging technologies such as augmented reality (AR) visualization of machine toolpaths to support clay 3D printing \cite{passananti2024augmented}.

Those previous works demonstrate growing interest in integrating ceramics with digital fabrication, highlighting opportunities to expand creative workflows. 

\subsection{Application of Generative AI in Fabrication}
Generative AI tools, powered by large language models (LLMs), have capability to transform natural language prompts into coherent outputs such as visual images or explanatory texts \cite{rombach_high-resolution_2022, mahdavi_goloujeh_is_2024}. The development of these tools has made computational methods more actively participate in the design and creative process \cite{elran_fine_2023}. Their applications include, but are not limited to design, art, and writing \cite{chiou_designing_2023, AI_TexttoImage_Generator, Transformer_Poetry_Generation}.

More recently, Generative AI tools have begun to incorporate 3D design and modeling capabilities \cite{lee_impact_2024, chen_memovis_2024}. They can now support sketch-based modeling \cite{meshyai, tripo3d}, a widely used technique in which 2D sketches are transformed into 3D models \cite{thiault_spineloft_2025}. This advancement gives GenAI the potential to serve not only as a visualization tool but also as a bridge between conceptual design and fabrication-ready models, thereby assisting users in CAD and CAM tasks. It also opens new possibilities for integrating GenAI with digital fabrication with traditional craft practices \cite{elran_probabilistic_2024}. Previous research has demonstrated how GenAI can automatically generate 3D models with tactile textures or surface ornamentation that can be transferred into fabrication processes \cite{Faraz_TactStyle_2025, tanaka_haptics_2023, Faruqi_style2fab_2023}. 

Within the context of Chinese crafts, GenAI has also been applied in several studies, including paper cutting and filigree \cite{wang_harmonycut_2025, tao_aifiligree_2025}. These works demonstrate the potential of GenAI to capture cultural motifs and workflows while enabling new hybrid practices. Moreover, GenAI models can be further trained or fine-tuned with aesthetic and cultural knowledge, making them more responsive to domain-specific traditions.

In a word, these related works show an opportunity of GenAI as a promising tool for cultural craft innovation. Its multimodal capacity to transform text or sketches into 3D models and textures positions it as a potential mediator between traditional aesthetics and digital fabrication to integrate clay 3D printing into Chinese ceramic practices.

\subsection{Hybrid Fabrication Approaches in HCI}
Over the past decades, HCI researchers have increasingly investigated how CAM and CAD can be integrated with digital fabrication \cite{Del_Engaging_2024}. Computational systems have been developed with the capability to dynamically control machine behaviors, allowing them to align with manual craft production by connecting to digitally fabricated tools \cite{Jacobs_computational_2025}. As a result, digital fabrication technologies have been positioned as a bridge between computational methods and the making of physical artifacts \cite{Jacobs_digital_2016, Hirsch_Nothing_2023}. In this way, hybrid fabrication approaches have become important in contemporary craft research, as they demonstrate how digital processes can be transformed into physical productions \cite{zoran_Hybrid_2013, zoran_Hybrid_2015, Devendorf_Beyond_2017}.

Hybrid digital craftsmanship emphasizes not only early HCI goals such as efficiency and usability, but also the creation of expressive media that sustain engagement and enable diverse outcomes \cite{Jacobs_digital_2016}. For example, programming-based design methods have been combined with digital fabrication to support expressive and flexible creative practices in object design and fashion design \cite{Jacob_Codeableobjects_2013}.

With advances in enabling technologies, some hybrid digital fabrication workflows have been designed to reduce entry barriers for beginners and to support novices in developing manual skills \cite{Turakhia_Reimaging_2022}. Researchers have also explored how insights from craft practitioners can be integrated into computational design to produce fabrication techniques that remain compatible with domain-specific practices \cite{Jacobs_computational_2025, Throwing_Moyer_2024, lu_craftsfabrication_2022}. Related studies include efforts discussed in Sections 2.2 and 2.3, such as combining 3D printing with traditional handcraft \cite{frost_sketchpath_2024, friedman-gerlicz_weaveslicer_2024, bourgault_coilcam_2023}, applying augmented reality (AR) to clay printing workflows \cite{passananti2024augmented}, and incorporating Generative AI into digital fabrication systems to extend traditional crafts \cite{elran_probabilistic_2024}.

These studies suggest that hybrid digital fabrication approaches are less about replacing established craft knowledge than about enabling new possibilities through collaboration between emerging technologies and human creators\cite{he_i_2025,liu_salt_2025}. In this context, we explore a hybrid workflow that integrates Generative AI and clay 3D printing to support the design, production, and decoration of Chinese ceramics.

\section{Methodology}\label{sec:p1Methods}
To design a hybrid fabrication workflow that integrates Generative AI and clay 3D printing into the ceramic production process, we conducted a two-phase study informed by co-design and research through design methodologies \cite{zimmerman_research_2007,gaver_what_2012, Co-Design, zamenopoulos2018co}. These approaches enabled us to iteratively shape the workflow through direct engagement with creative practices. 



\subsection{Phase 1: Collaborative Design Study}
\begin{figure*}[htbp]
 \centering
 \includegraphics[width=1\textwidth]{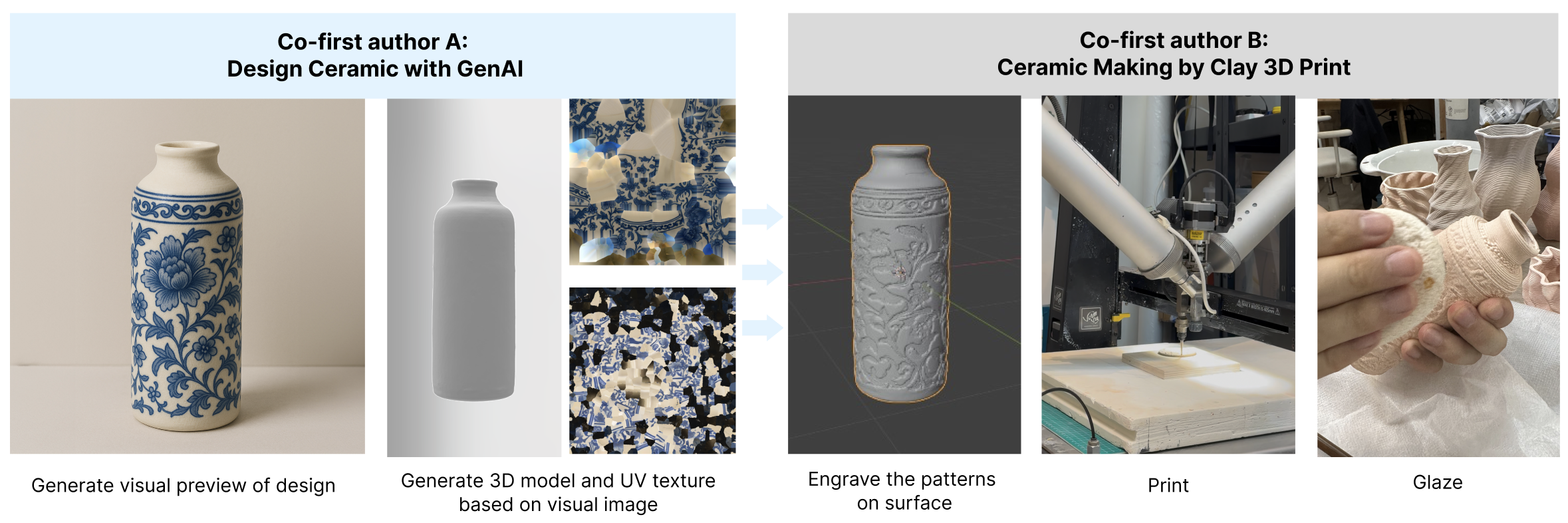}
 \caption{Phase 1: Collaborative Creation Study. Author A worked with GenAI to get ceramic design; Author B did clay print and glaze jobs.}
 \label{phase1}
\end{figure*}

Phase 1 involved a collaborative design process between the two co-first authors to explore a hybrid-making workflow. Co-first author A (Author A) is a researcher specializing in GenAI workflows for creative support, with extensive experience using GenAI tools across creative projects and familiarity with different model capabilities. Co-first author B (Author B) is a ceramic artist and researcher specializing in clay 3D printing, with expertise in both traditional handcraft techniques and hybrid digital fabrication. In the first phase, the two researchers began by learning from each other’s expertise through a collaborative creation process. After exchanging knowledge of GenAI and clay 3D printing, they co-designed a creative workflow and applied it to produce ceramic works (Fig.\ref{phase1}).

Text-to-image multimodal LLMs were applied for this workflow because their conversational prompting allowed iterative refinement \cite{openai_introducing_nodate}, which was necessary for tuning outputs toward Chinese ceramic aesthetics. As prompts required continual adjustment using visual references and cues, Author A led this stage due to her experience in developing such prompting strategies. The resulting previews were then transformed into 3D forms using Tripo and Meshy, which produced 3D models and texture maps for further adaptation. Author B subsequently translated these models into fabrication-ready geometry using Grasshopper, preparing them for clay 3D printing. After printing and firing, glazing trials were conducted to complete the ceramic piece.

Throughout the process, the two researchers continuously exchanged expertise and provided feedback to one another. Their final reflections and evaluations informed the design of the second phase of the study.

%



\subsection{Phase 2: Empirical Study}
To evaluate the workflow developed in Phase 1, we conducted a second phase of the study to observe how craft creators engaged with the proposed process. Inspired by previous HCI and design research that involved craft practitioners in examining practical innovations and creative challenges \cite{toka_practice-driven_2024, Devendorf_Craftspeople_2020, muller_participatory_2002}, we invited four participating creators, two ceramic artists and two enthusiastic beginners to engage in the study. We introduced ClayScape, the tool developed after Phase 1, in this phase as a convenient way for participants to engage with and test the proposed workflow. The tool included prompt workspace for generating design previews and textured 3D models, along with a print preview area that allowed participants to visualize how their creations would appear when fabricated in clay. We clarify that this study should be understood as an exploratory investigation into the workflow’s potential, rather than a definitive validation of ClayScape as an independent system.

This phase was divided into two sessions because the printed ceramic works required a drying period before they could be glazed. The first was a guided design session, where Author A guided the participating creators through the use of ClayScape to generate design previews and AI-generated 3D models. During this creative process, they were encouraged to think aloud, verbalizing their decisions and reactions. Once the designs and models were iteratively finalized, they were prepared and printed. Author B intervened in the G-code during the printing process when necessary to ensure the feasibility of the fabricated forms. In the second session, held after the prints had dried, the participating creators were invited to attend in person at Author B's ceramic studio. Author B introduced the printed clay objects and invited each creator to decorate them based on the engraved surface textures. Participants were encouraged to use their preferred techniques for glazing or painting. In this phase, the researchers primarily observed and provided clarification on tool usage and material constraints when needed, while avoiding direct intervention in participants’ creative decisions.

\begin{figure*}[htbp]
 \centering
 \includegraphics[width=1\textwidth]{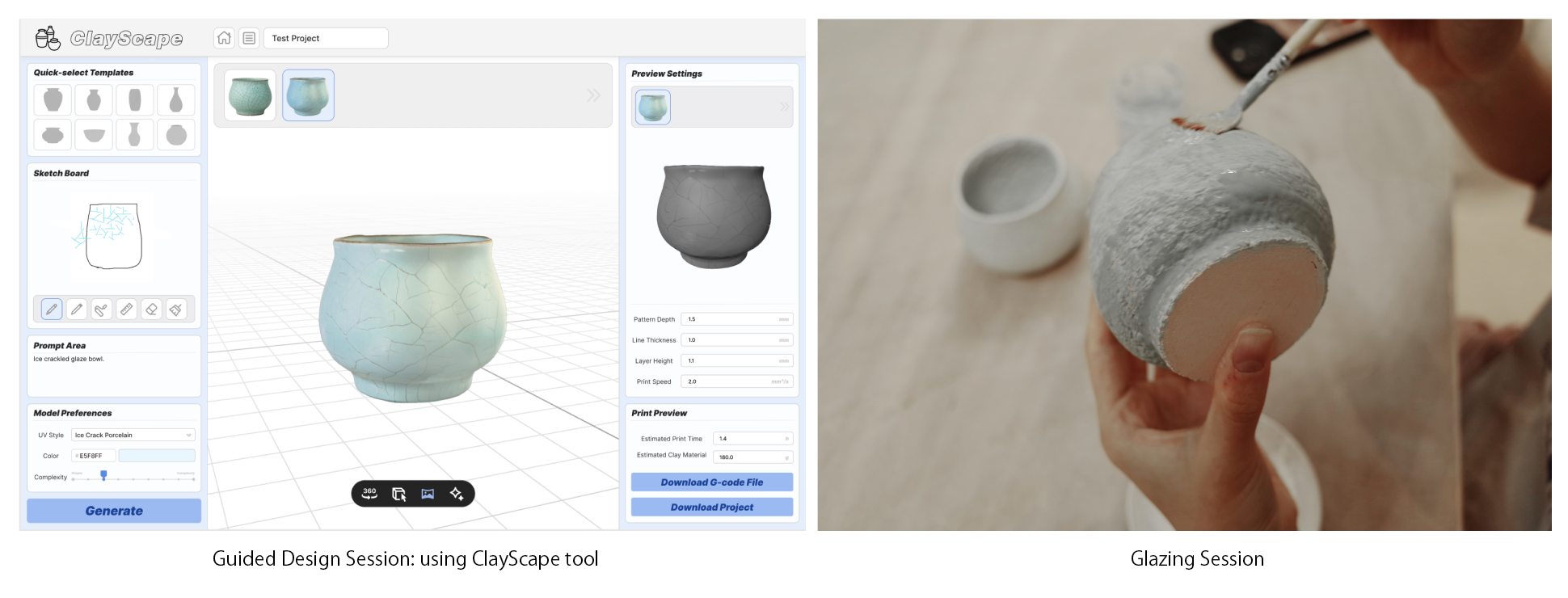}
 \caption{Two sessions of Phase 2 Empirical Study. A guided design session: Participating creators were guided to use ClayScape tool for ceramic design (Left). Glazing session at Author B's ceramic studio (Right)}
 \label{phase2method}
\end{figure*}

We conducted semi-structured interviews (approximately 30 minutes) after each session. The interviews mainly focused on capturing the participating creators’ reflections on the experience, including their attitudes toward the integration of GenAI and 3D printing into craft practice, comparisons with traditional ceramic-making methods, and the challenges they encountered throughout the workflow.

\subsection{Participating Craft Creators}
We used convenience sampling to recruit four creators with varying levels of ceramic experience (Table.\ref{table1}). Author B invited two ceramic artists through his professional network, while Author A invited two ceramic enthusiasts who had previously attended a ceramic training program with her. All of the participating creators were Chinese and familiar with the cultural context of Chinese ceramic aesthetics.

While all participants reported some prior exposure to GenAI tools, this experience was primarily limited to early-stage ideation or visual inspiration using text-to-image systems, rather than structured prompt engineering or material-aware generation. Prior 3D printing experience also varied and was generally not specific to clay-based fabrication; while present, it involved basic digital fabrication workflows such as file preparation or printer operation. None of the participants had previously combined GenAI with ceramic creation or clay 3D printing.

The variation in participants’ ceramic, AI, and fabrication backgrounds allowed us to observe how creators with different levels of experience engaged with the workflow and to evaluate how it supported diverse creative approaches.

\begin{table*}[h]
\begin{tabular}{l l l l l}
\toprule
\textbf{ID}  & \textbf{Professions} & \textbf{Ceramic Experience} &\textbf{3D Print Experience} &\textbf{GenAI Experience}\\ 
\midrule
A01 & Graduate Student  & Beginner (less than 1 year) & Yes & Yes \\ 
A02 & Product Designer  &  Beginner (less than 1 year) & No & Yes  \\ 
A03 &  Pottery Tutor & Advanced (5+ years) & No & Yes \\
A04 &  Assistant Professor & Advanced (5+ years) & Yes & Yes  \\
\bottomrule
\end{tabular}
  \caption{Demographics of Participating Craft Creators}
  \label{table1}
\end{table*}

All participating craft creators provided informed consent for the collection and use of photos, videos, text records of their creations, interview data, and workshop interactions in this study.

\subsection{Data Analysis}
In this phase, we adopted a reflective research through design approach to collect data \cite{Dalsgaard_Reflective_2012, Nimkulrat_hands_2012}. Each co-first author documented their own process in detail, including the use of generative prompts, AI output selection, 3D modeling adjustments, engraving methods, and glazing guided by surface textures. These records served as both internal feedback for shaping the ClayScape prototype and as qualitative data for the whole investigation.

In the second phase, we collected the observation notes, think-aloud data, and interview transcripts during the two sessions for analysis. The co-first authors conducted an inductive thematic analysis of all the data to summarize participants' reflections, creative approaches, and main challenges during the workshops to obtain qualitative codes \cite{braun_using_2006, thomas_general_2006, wicks_coding_2017}.

The coding process was iterative. The two researchers began by re-reading the transcripts of the interviews and think-aloud notes independently to deepen their understanding of the content. They then applied open coding to identify initial codes in the interviews, observations, and think-aloud notes that captured key ideas, meanings, or concepts relevant to addressing the research gap \cite{braun_using_2006}. After the initial coding phase, the researchers collaborated to identify further themes from the generated codes. They continually refined the codes and classified them into themes over several rounds until they reached an agreement that the themes accurately represented the data.

\section{Phase 1: Workflow Exploration}\label{sec:p1 findings}
\subsection{Knowledge Exchange and Workflow Framing}
\subsubsection{Exploring GenAI for Ceramic Design}

At the beginning of Phase 1, Author A introduced the capabilities of current GenAI tools that can generate design images and 3D models, giving Author B a clear sense of how AI could support CAD and CAM. In turn, Author B presented the traditional ceramic-making process and the current developments in clay printing. After exchanging expertise, both authors collaboratively designed a hybrid fabrication workflow, where GenAI models supported early design exploration and human creators guided fabrication decisions from digital modeling to clay 3D printing to realize a physical ceramic piece.

\begin{figure*}[h]
 \centering
 \includegraphics[width=0.8\textwidth]{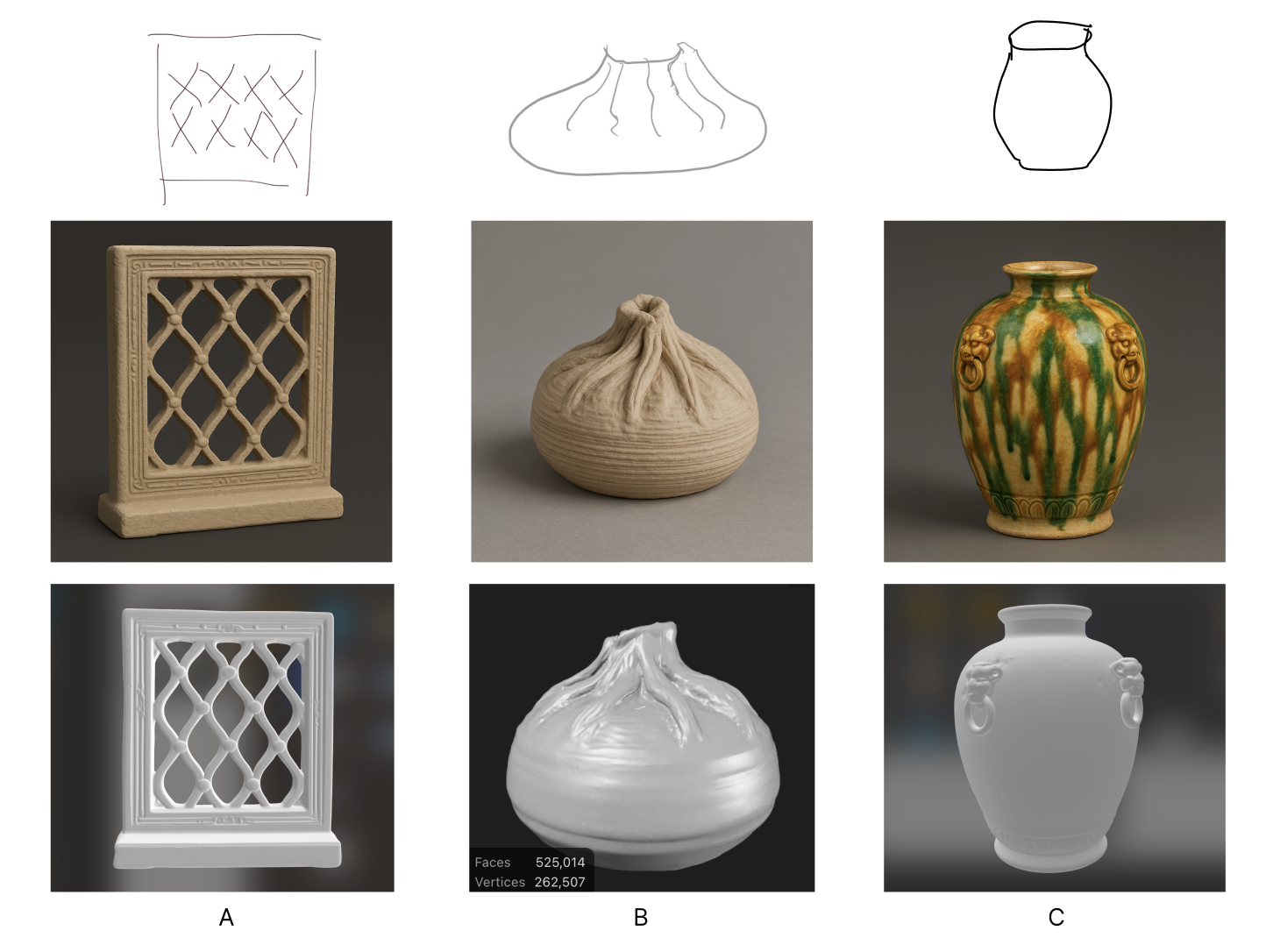}
 \caption{Author A presented three AI-collaborated designs to Author B: (A) Lattice-structured ceramic screen with Chinese motifs; (B) Baozi-shaped sculpture; (C) Tang Sancai jar with dragon relief}
 \label{phase 1 test}
\end{figure*}

To explore the potential of GenAI in this workflow, Author A conducted a series of trial generations using text-to-image LLMs to produce images of textured Chinese ceramic forms. She employed minimal line sketches and concise textual descriptions as prompts, reflecting common early-stage practices in ceramic design. These generated images were then treated as visual design references and fed into 3D generative tools to generate corresponding models and UV textures. Author A presented three designs to Author B: a lattice-structured ceramic screen with Chinese motifs, a baozi-shaped sculpture, and a Tang Sancai jar with dragon relief (Fig.\ref{phase 1 test}).

\subsubsection{Author B: incorporating GenAI into ceramic making process}
After reviewing the AI-generated results, Author B reflected that the generated forms and motifs aligned with established Chinese ceramic aesthetics, and the visual and 3D outputs were sufficient for fabrication guidance. It led him to think about how Generative AI could potentially support certain steps in traditional ceramic workflows.

Based on his expertise and working experience, Author B noted that sketching is typically the first step in ceramic craft practice. Besides as final appearance previews, design sketches could be as guides for shaping and fabrication decisions before physical making. However, hand sketches often remain as simple outlines of shapes, as producing detailed drawings with surface patterns and colors can be time-consuming. Even when more complete design drawings are needed, craftspeople often delay this step to periods of waiting, such as during drying or kiln firing.  Although existing computational rendering tools can visualize sketches into more detailed representations, modeling and rendering are not necessarily time-efficient and often require CAD skills that present a high entry barrier for many ceramic practitioners. After observing the AI-generated designs, Author B recognized that Generative AI could enhance this stage by producing realistic renderings directly from simple sketches and text prompts. This approach not only reduces the time required for design visualization but also provides a more accessible entry point for creators who are less confident in detailed drawing.

Throwing and molding are core stages in traditional ceramic making, as they directly determine vessel form. However, making complex geometries and fine structural details often requires substantial skill and repeated practice even for experienced ceramic artists. Although computational 3D modeling and clay printing provide alternative ways to support form-making, particularly for achieving complex or repeatable shapes, these digital approaches introduce technical barriers. Not all ceramic artists are skilled in CAD or CAM, and even for experienced practitioners, building high-polycount models is time-intensive. In this context, Author B saw value in 3D GenAI tools as a way to generate base forms more efficiently, lowering the entry barrier to clay printing while still allowing artists to focus on material execution and aesthetic decisions.

Decoration and glazing in ceramic practice typically require freehand painting skills and accumulated experiences. Author B envisioned that AI-generated textures could be engraved onto ceramic surfaces to serve as visual and tactile guides, with patterns iteratively adjusted during the generation stage. This approach allows decorative decisions to be shaped before fabrication, lowering the barrier to decoration while preserving space for individual aesthetic choices.

\begin{figure*}[h]
 \centering
 \includegraphics[width=0.8\textwidth]{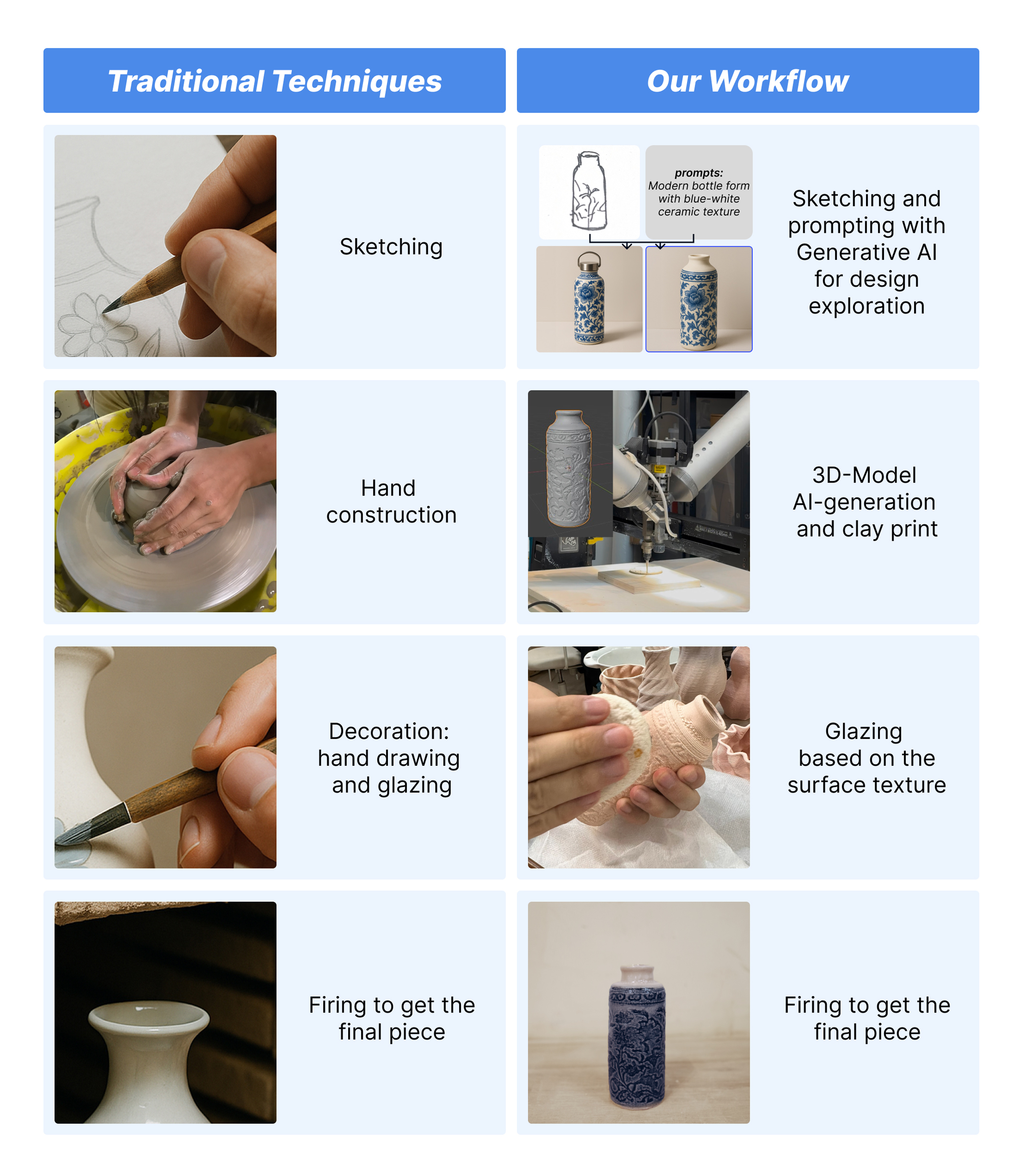}
 \caption{Correspondence between traditional ceramic-making techniques and our proposed workflow. Each row illustrates one stage, showing how traditional processes such as sketching, hand construction, decoration, and firing are paralleled by AI-supported design, 3D clay printing, texture-guided glazing, and final firing in our workflow.}
 \label{step comparison}
\end{figure*}

Based on these considerations, we designed a hybrid workflow that lowers barriers to ceramic fabrication by following the conventional sequence of ceramic processes while integrating Generative AI and clay 3D printing. (Fig.\ref{step comparison}).

\subsection{Creation Trials}
\subsubsection{Author A: Co-created Designs with GenAI}
From early explorations, Author A observed that progressively introducing visual references enabled the LLMs to produce outputs more closely aligned with intended culturally aesthetic expectations. In this phase, the model was guided within Chinese ceramic visual references selected by Author B, including Qinghua brush motifs, black-and-white ink patterns, and Tang sancai glaze palettes, drawn from established ceramic scholarship \cite{pierson2009chinese}. These categories represent recurring taxonomies in Chinese ceramic history and ensured that the generated outputs reflected core components of Chinese aesthetics, where vessel form, symbolic motifs, and glaze/color language are closely integrated.

Beyond generating conventional ceramic forms, Author A explored hybrid outputs that combined unconventional geometries with Chinese decorative patterns, as well as traditional forms rendered through contemporary textures. This exploration aligns with broader movements in contemporary Chinese craft that seek to move beyond traditional vessel typologies and reinterpret cultural aesthetics for modern contexts \cite{liu_craft_2024, jingexamining}.

\begin{figure*}[h]
    \centering
    \begin{subfigure}{0.19\textwidth}
        \centering
        \includegraphics[width=\linewidth]{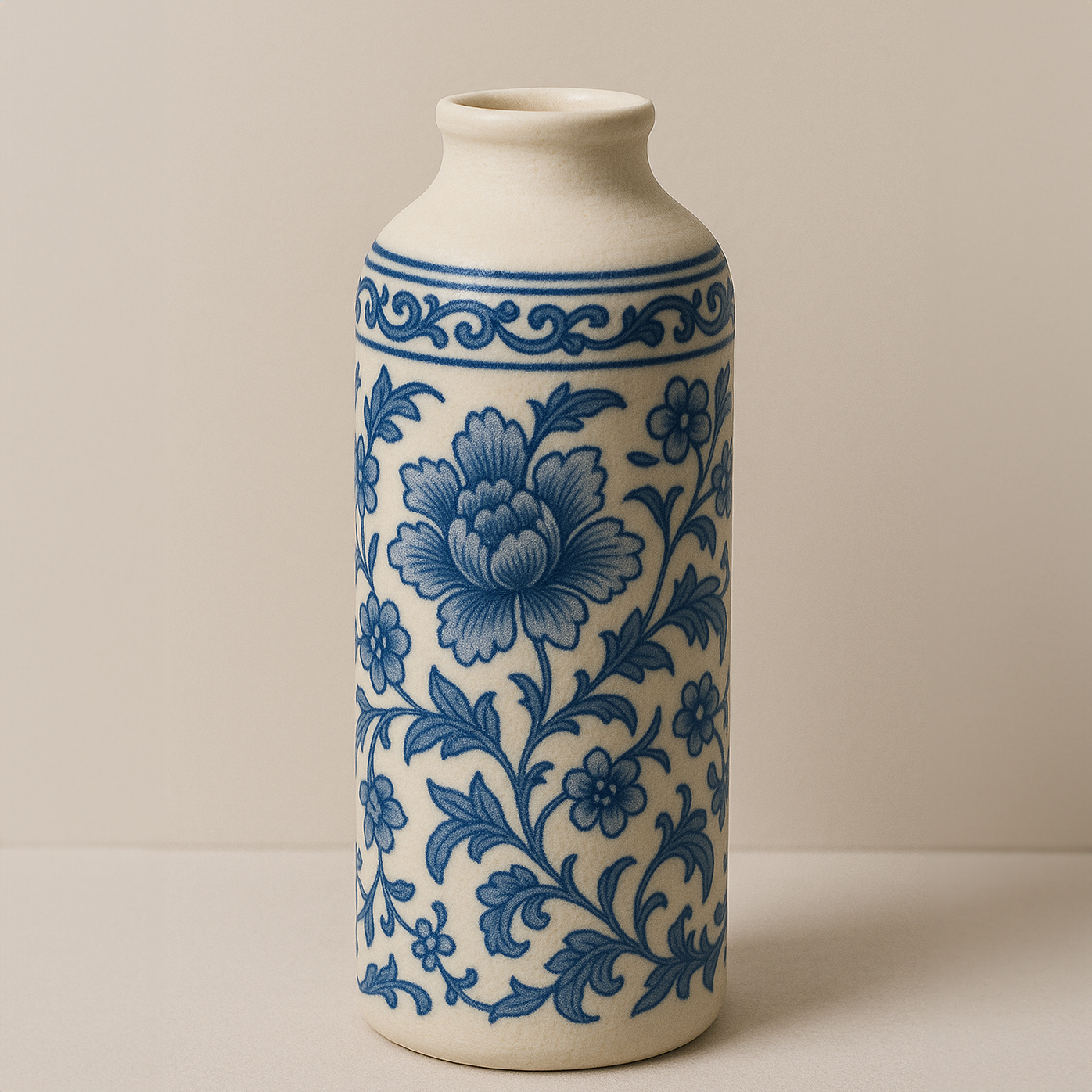}
        \caption{}
        \label{fig:qinghua}
    \end{subfigure}
    \begin{subfigure}{0.19\textwidth}
        \centering
        \includegraphics[width=\linewidth]{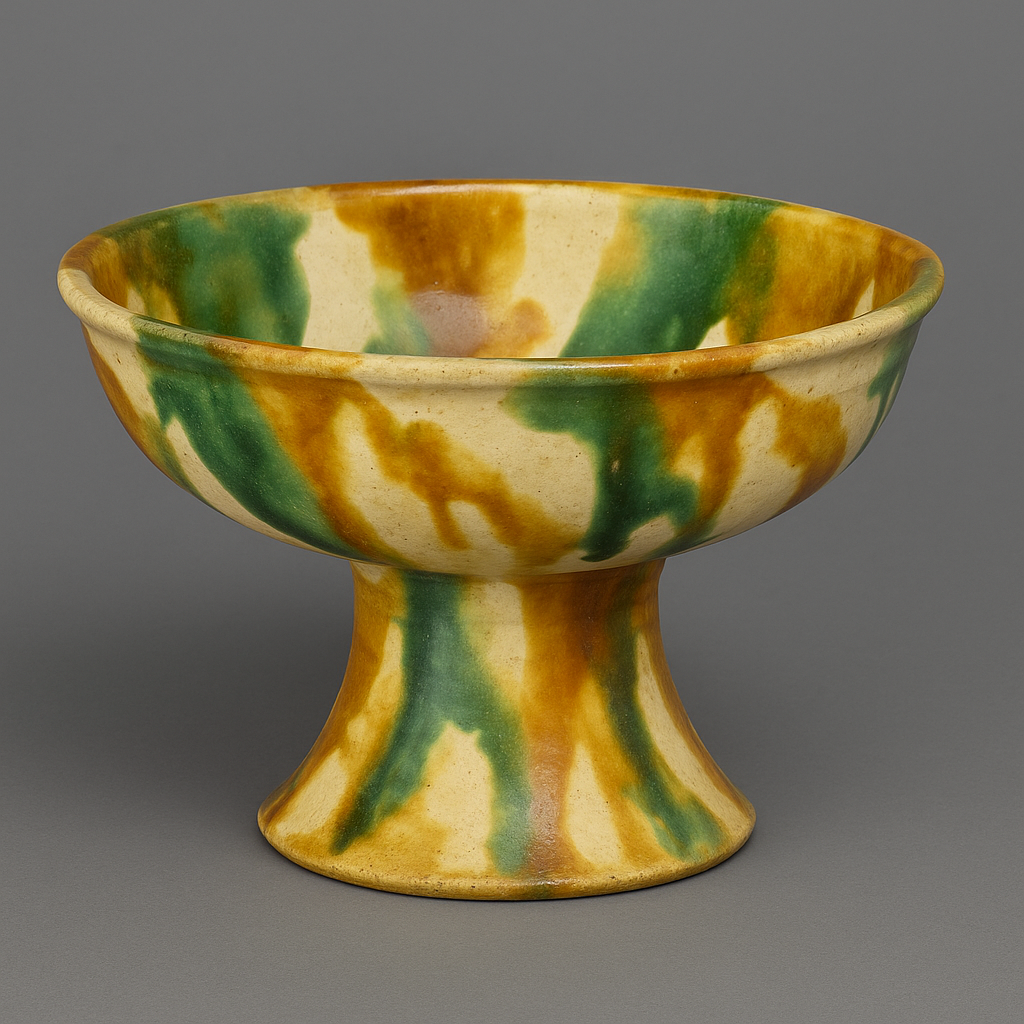}
        \caption{}
        \label{fig:tangsancai}
    \end{subfigure}%
    \begin{subfigure}{0.19\textwidth}
        \centering
        \includegraphics[width=\linewidth]{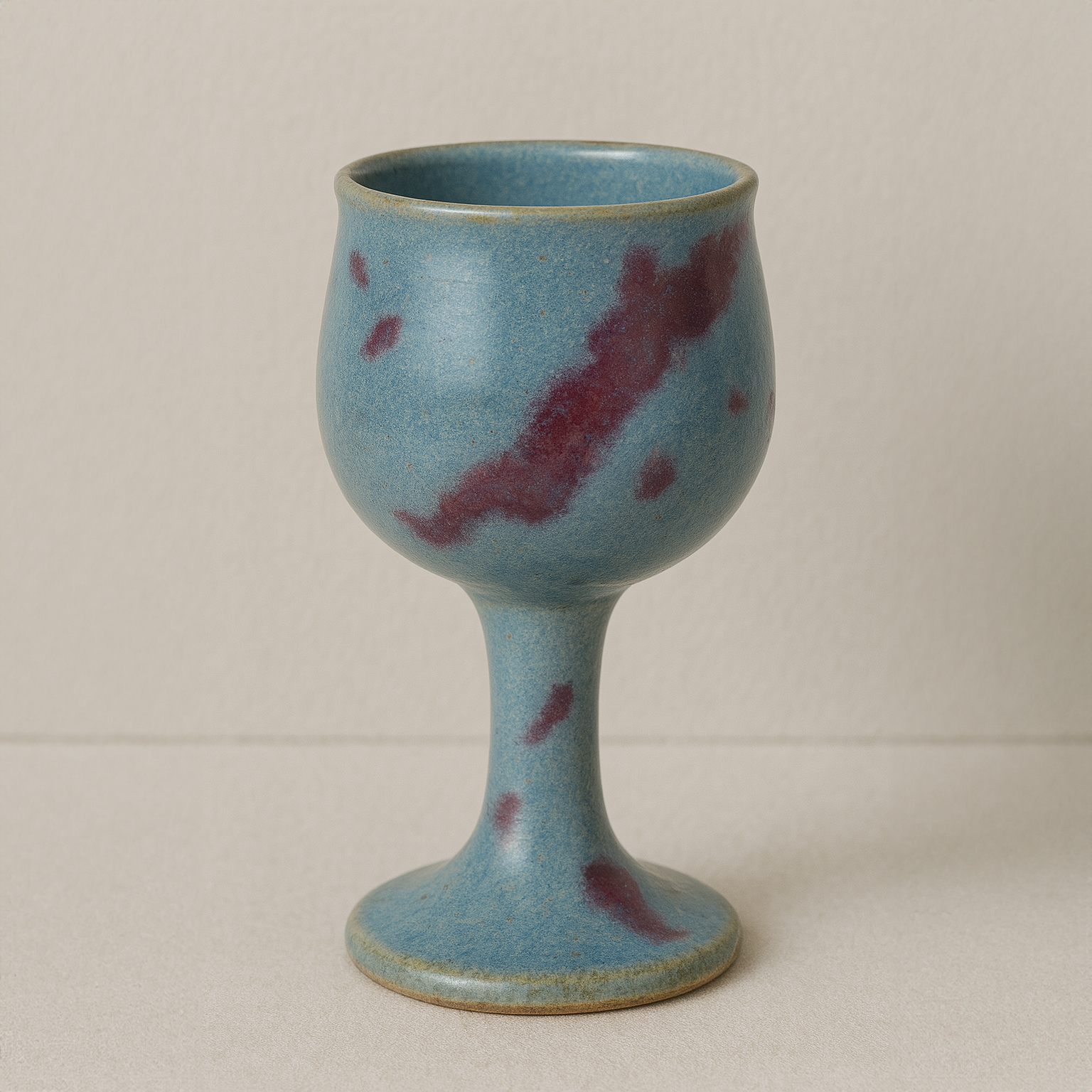}
        \caption{}
        \label{fig:bluechalice}
    \end{subfigure}%
    \begin{subfigure}{0.19\textwidth}
        \centering
        \includegraphics[width=\linewidth]{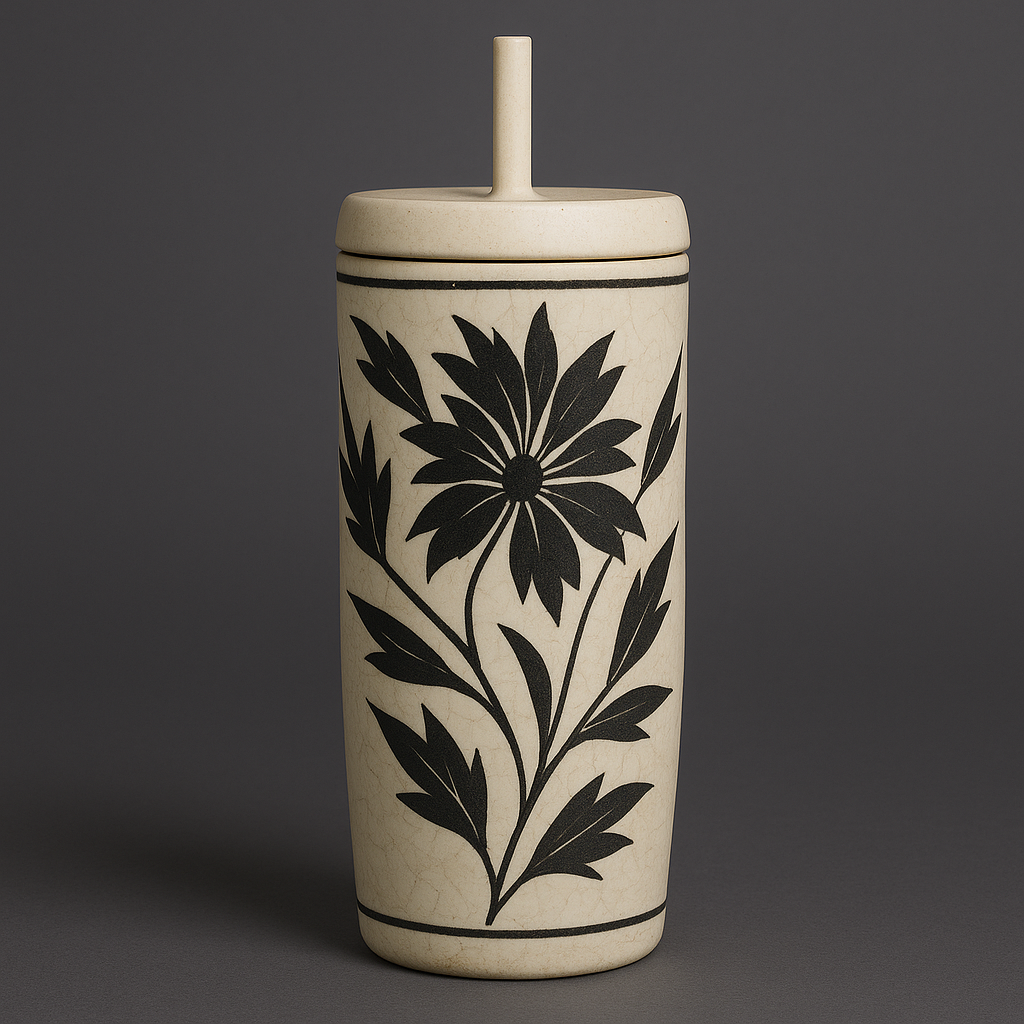}
        \caption{}
        \label{fig:strawbottle}
    \end{subfigure}%
    \begin{subfigure}{0.19\textwidth}
        \centering
        \includegraphics[width=\linewidth]{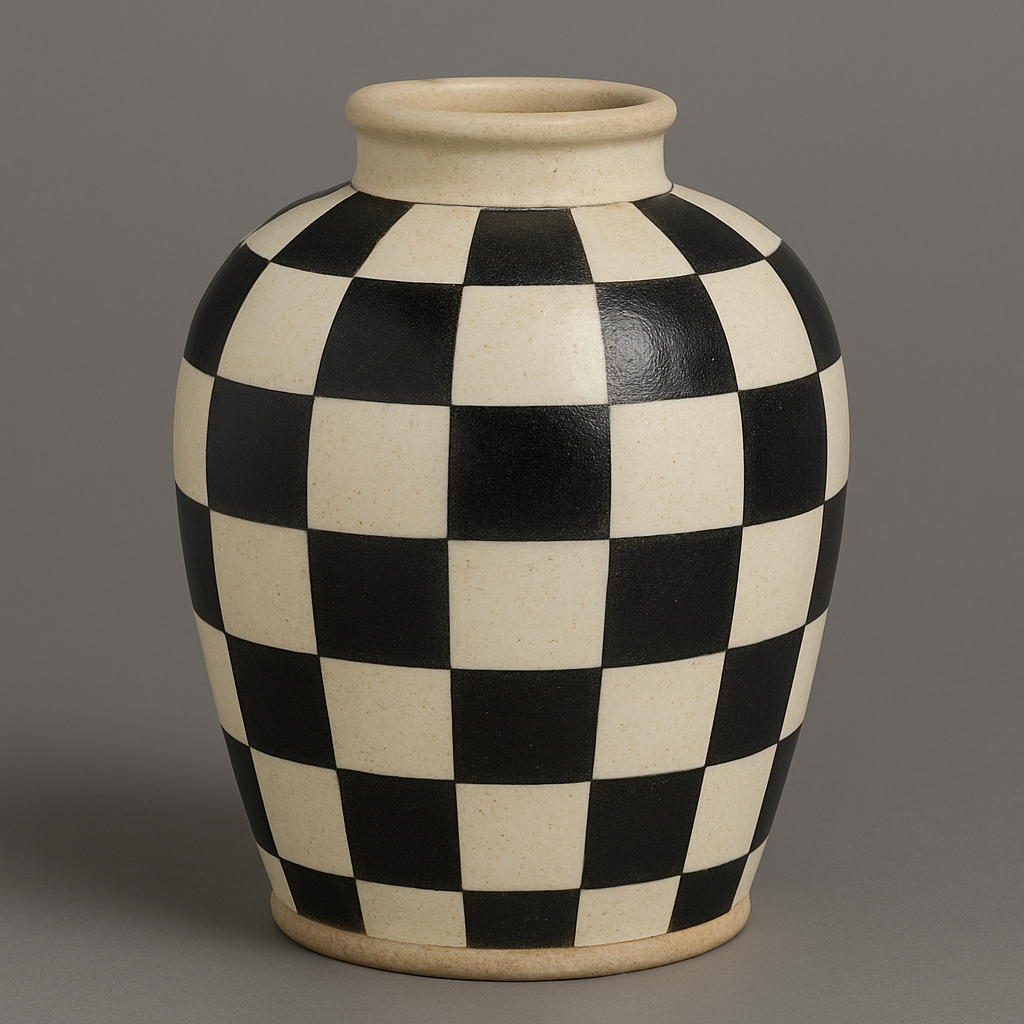}
        \caption{}
        \label{fig:checkboardjar}
    \end{subfigure}
    \caption{(a): Qing Hua Vase: a traditional form with classic Qing Hua blue-and-white patterns; 
    (b): Tang Sancai Bowl: a traditional shape with sancai glazing; 
    (c): Blue Glaze Chalice: a contemporary glass-inspired form with traditional Chinese glaze colors; 
    (d): Straw Bottle with Black Floral Patterns: a modern bottle with black-and-white floral motifs; 
    (e): Checkerboard Jar: a traditional jar shape combined with a modern checkerboard pattern.}
    \label{fig:generated_design}
\end{figure*}

Author A finally produced five designs using GenAI tools:

\begin{itemize}
    \item \textbf{Qing Hua Vase}: a traditional form with classic Qing Hua blue-and-white patterns (Fig.\ref{fig:qinghua}).
\end{itemize}
\begin{itemize}
    \item \textbf{Tang Sancai Bowl}: a traditional bowl shape with Tang Sancai glazing (Fig.\ref{fig:tangsancai}).
\end{itemize}
\begin{itemize}
    \item \textbf{Blue Glaze Chalice}: a contemporary glass-inspired form with traditional Chinese glaze colors (Fig.\ref{fig:bluechalice}).
\end{itemize}
\begin{itemize}
    \item \textbf{Straw Bottle with Black Floral Patterns}: a modern bottle with black-and-white floral motifs (Fig.\ref{fig:strawbottle}).
\end{itemize}
\begin{itemize}
    \item \textbf{Checkerboard Jar}: a traditional jar shape combined with a modern checkerboard pattern (Fig.\ref{fig:checkboardjar}).
\end{itemize}

\subsubsection{Author B: Engraving Surfaces and Clay Printing}
\begin{figure*}[h]
 \centering
 \includegraphics[width=0.8\textwidth]{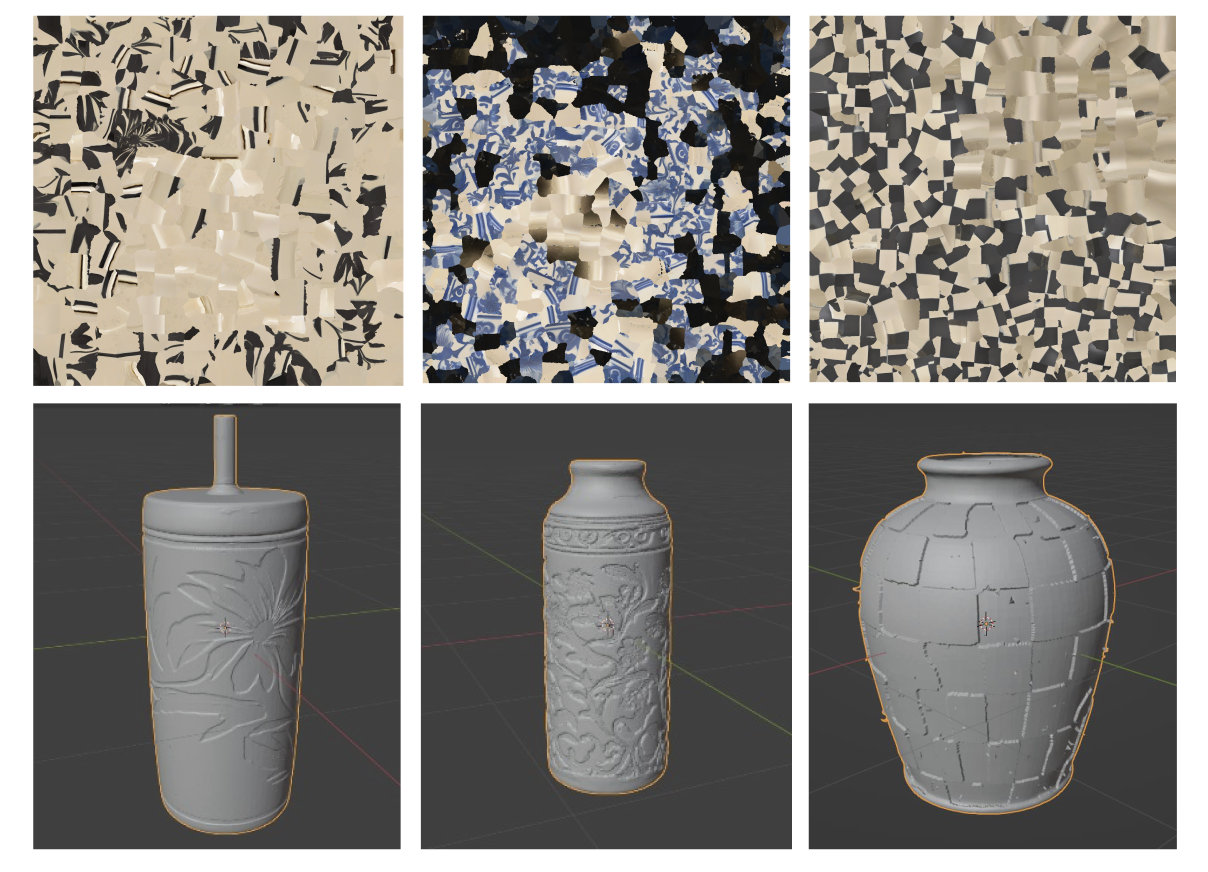}
 \caption{AI-generated UV textures (top); Author B engraved the patterns on the surfaces of AI-generated models (bottom)}
 \label{engravaing}
\end{figure*}

Once the models were received, Author B imported them into Rhino and Grasshopper to engrave patterns onto the surfaces. This stage required balancing visual clarity with material constraints, as the engraved motifs needed to remain legible after firing while supporting subsequent glazing. As a result, engraving depth was iteratively adjusted to accommodate these fabrication considerations (Fig.\ref{engravaing}).

\begin{figure*}[h]
 \centering
 \includegraphics[width=0.8\textwidth]{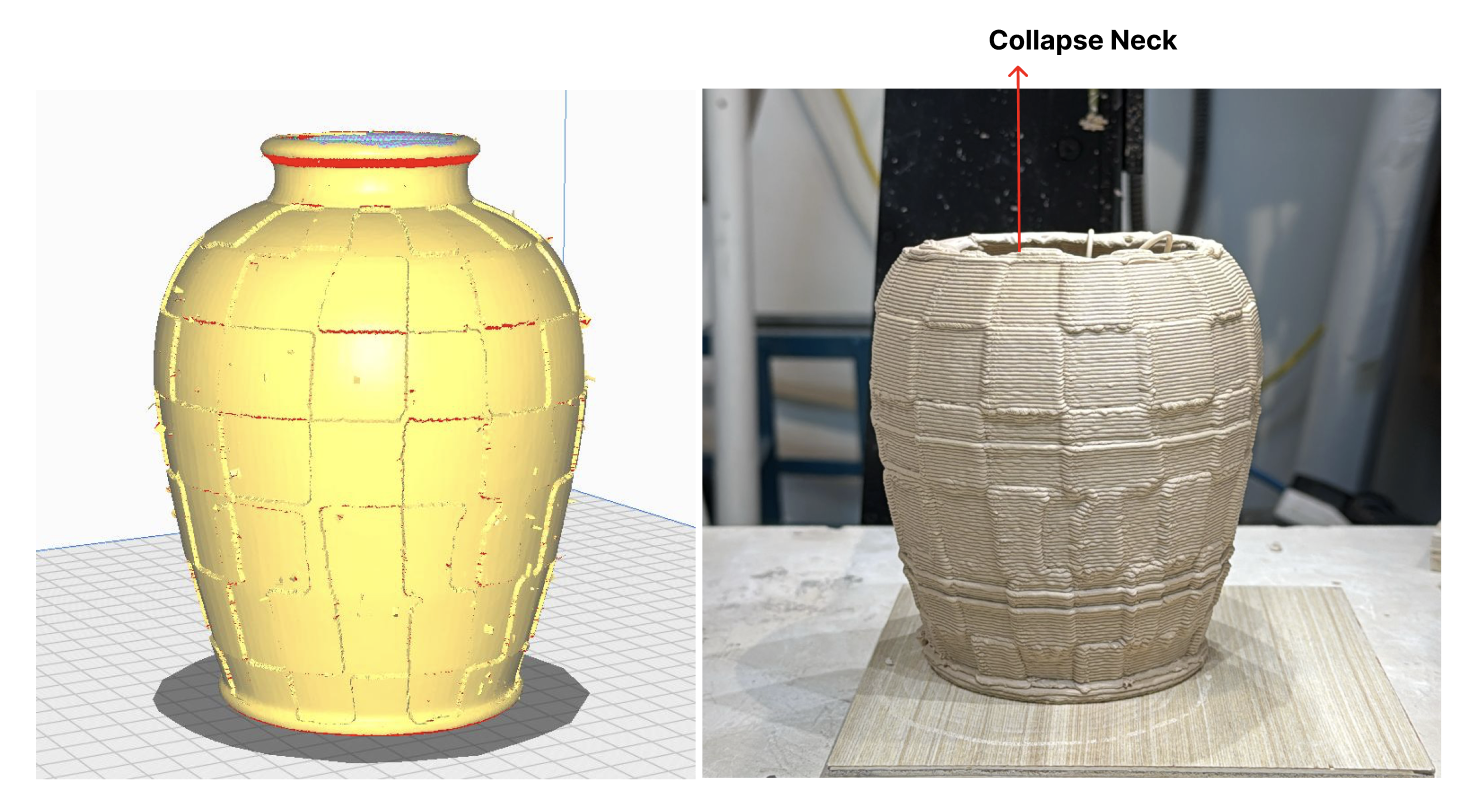}
 \caption{Structural analysis in Grasshopper (left) highlights high-risk areas of the generated model in red. The corresponding clay print (right) confirms this prediction, with the vessel’s neck collapsing during the printing process.}
 \label{collapse}
\end{figure*}

During the clay printing process, several issues were found. The most critical was printability: steep overhangs and thin structures often collapsed in clay, even when Grasshopper simulations predicted failure in advance. An example of Checkboard Jar in (Fig.\ref{collapse}), where the Grasshopper preview highlights the neck area as structurally risky (red part), and the actual print subsequently collapsed at the same location. As a result, not all AI-generated models could be successfully printed.

After multiple trials, Author B managed to print the Qing Hua vase design for glazing tests. Two versions were produced with different engraving depths to determine the minimum effective depth for glazing guides. Author B used a basic glaze inlay technique by filling glaze into the dented texture and removing excess glaze with sponge, both vases were successfully finished.

\begin{figure*}[htbp]
    \centering
    \begin{subfigure}{0.25\textwidth}
        \centering
        \includegraphics[width=\linewidth, height=4cm]{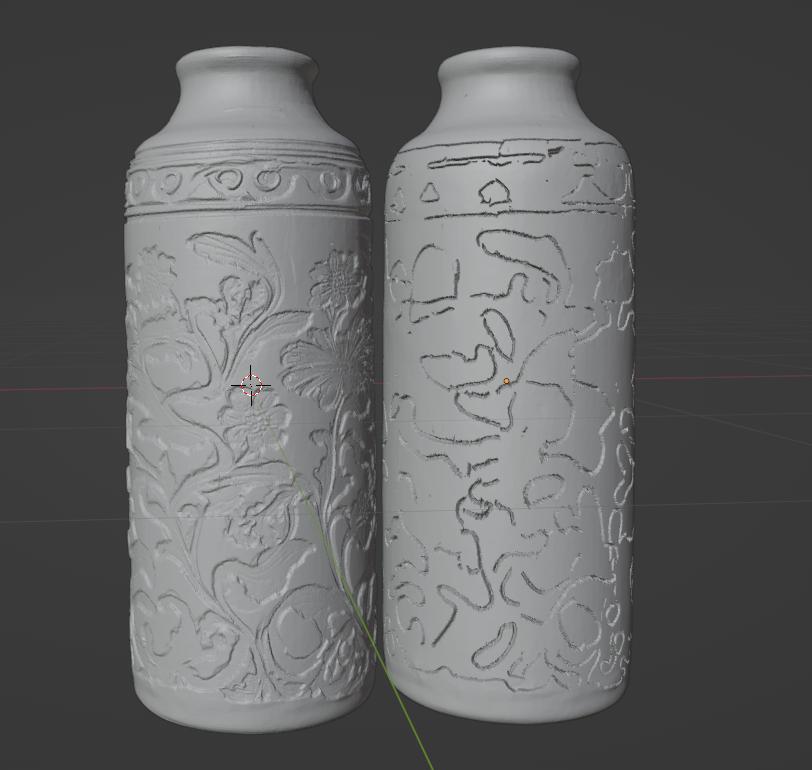}
        \caption{}
        \label{fig:bottlesengraving}
    \end{subfigure}
    \begin{subfigure}{0.25\textwidth}
        \centering
        \includegraphics[width=\linewidth, height=4cm]{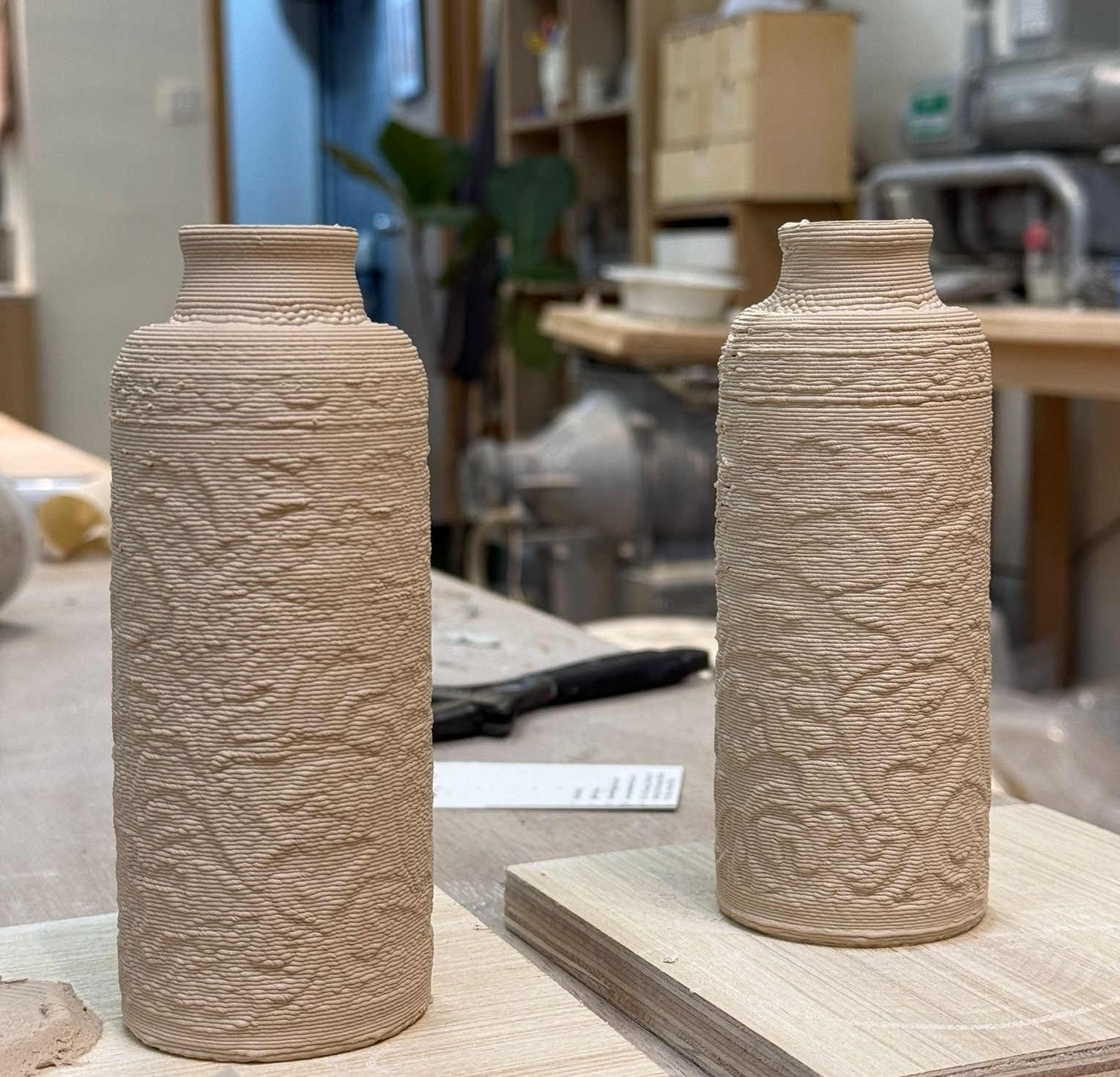}
        \caption{}
        \label{fig:printedbottles}
    \end{subfigure}%
    \begin{subfigure}{0.25\textwidth}
        \centering
        \includegraphics[width=\linewidth, height=4cm]{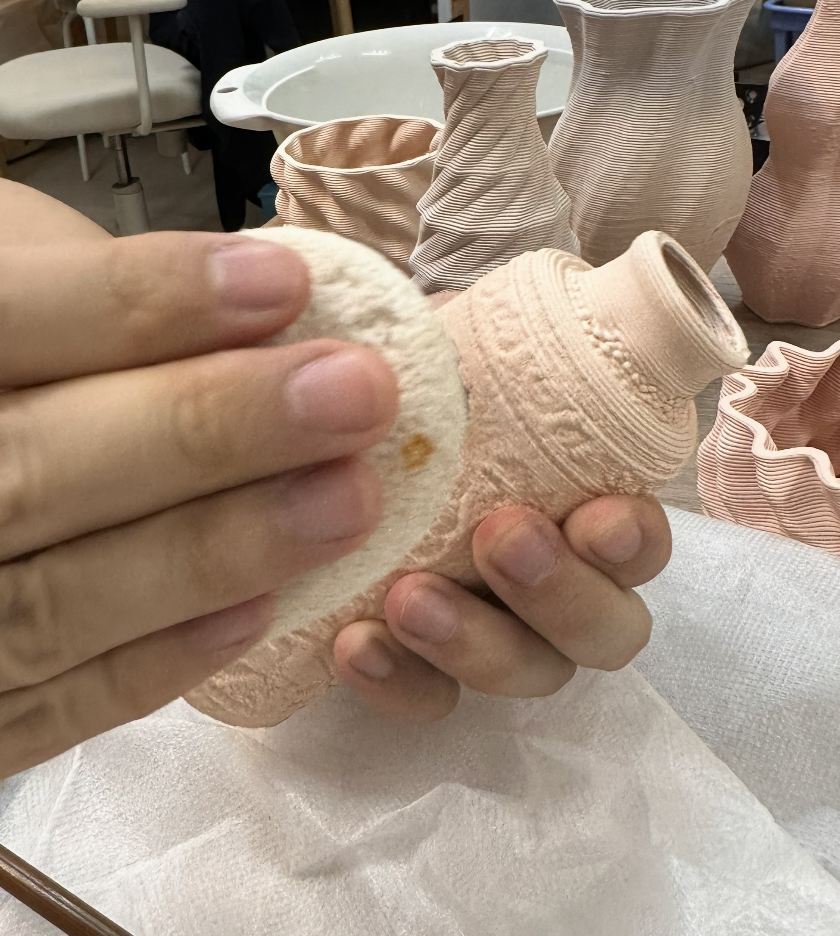}
        \caption{}
        \label{fig:spongeglaze}
    \end{subfigure}%
    \begin{subfigure}{0.25\textwidth}
        \centering
        \includegraphics[width=\linewidth, height=4cm]{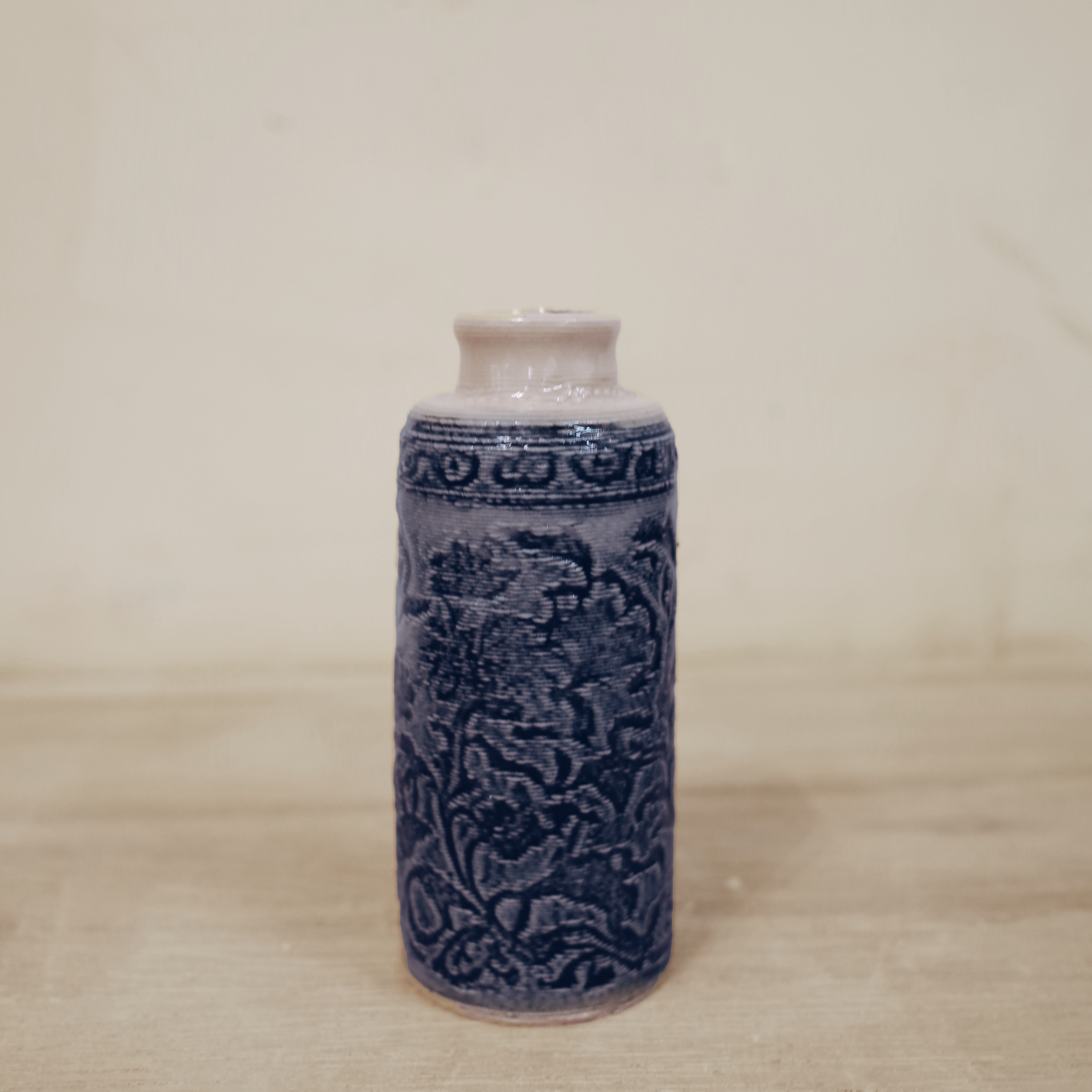}
        \caption{}
        \label{fig:final}
    \end{subfigure}%
    
    \caption{(a) Author B engraved the surface patterns of the Qing Hua vase with different depths; (b) 3D clay-printed versions produced with varying engraving depths; (c) glazing trial using sponge application; (d) final glazed piece showing visible surface patterns.}
    \label{fig:printed trial}
\end{figure*}

The final pieces highlighted two main challenges. First, structural instability led to collapses in certain parts of the prints. Second, the generated patterns were sometimes irregular, with vague lines that made it difficult to engrave clean details.

\subsection{Insights for Workflow Development}
Both researchers reflected on the trials and suggested improvements for the workflow and future tool development.

From Author A’s perspective, as an artist and designer with only entry-level ceramic experience, the AI-generated designs were inspiring but difficult to realize. She considered clay print a promising method for enabling artists like herself to create ceramic works. However, she also noted that inexperienced users may not obtain satisfactory GenAI outputs without clear prompting strategies.

\textbf{Author A’s suggestions}:

\begin{enumerate}
    \item Provide both sketches and visual references (shapes, patterns) alongside text prompts, to guide GenAI effectively.
\end{enumerate}

\begin{enumerate}
    \item Offer guidance on clay print constraints (e.g., dangerous shapes or angles) for users without printing knowledge.
\end{enumerate}

\begin{enumerate}
    \item Ensure UV textures have clean lines to support clear surface decoration.
\end{enumerate}

From Author B’s perspective, based on his prior experience transforming hand sketches into engraved 3D models, GenAI clearly improved efficiency. Still, some non-reducible steps remain, particularly structural analysis for printability.

\textbf{Author B’s suggestions}:

\begin{enumerate}
    \item Avoid steep angles and thin structures; keep slopes under ~45°.
\end{enumerate}

\begin{enumerate}
    \item Engraving depth of 1–1.5 mm is optimal for guiding glaze, requiring consideration of wall thickness.
\end{enumerate}

\begin{enumerate}
    \item Patterns with clear outlines work best for surface decoration.
\end{enumerate}

These reflections were passed on to the third collaborating researcher, who collaborated to design a system tool to further evaluate and refine the workflow in the next phase of the empirical study.

\section{ClayScape Tool Design}\label{sec:p2 method}
Drawing on insights from Phase 1, we developed an GenAI-driven tool ClayScape to make the workflow more accessible for creators and to support their engagement with the design process. It served not as a prototype for testing, but as an optimized GenAI-supported design tool, enabling artists to experience and contribute to the evolving workflow.

\begin{figure*}[htbp]
 \centering
 \includegraphics[width=1\textwidth]{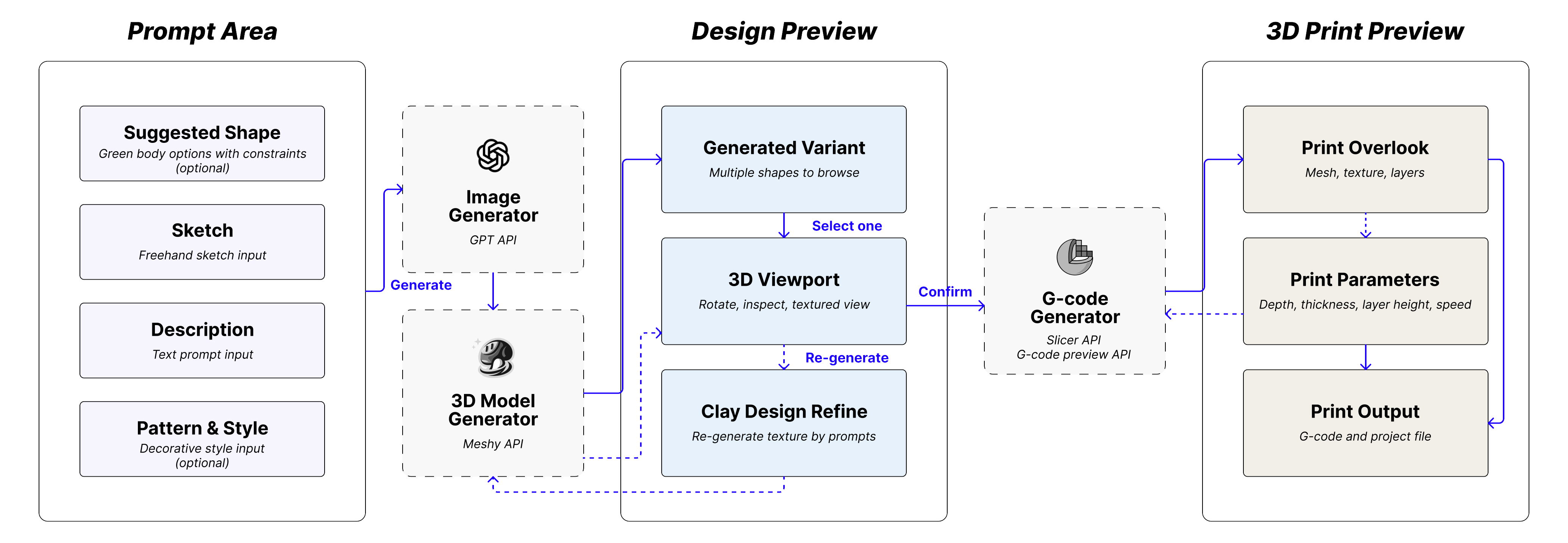}
 \caption{Architecture diagram of ClayScape, showing the underlying generative pipeline and data flow that support the interface presented in Figure \ref{interface}. The flow transforms multimodal user inputs into 3D models via integrated LLM and 3D generation APIs, enables iterative refinement through preview and re-generation, and produces fabrication-ready G-code with configurable printing parameters.}
 \label{architecturediagram}
\end{figure*}

ClayScape follows a generative pipeline that transforms multimodal user inputs into fabrication-ready outputs (Fig.~\ref{architecturediagram}). User inputs, including sketches, textual descriptions, and optional stylistic constraints, are processed through integrated LLM-based image generation and 3D model generation services. The system supports iterative refinement through preview and re-generation, before producing G-code with configurable printing parameters. 

\begin{figure*}[htbp]
 \centering
 \includegraphics[width=1\textwidth]{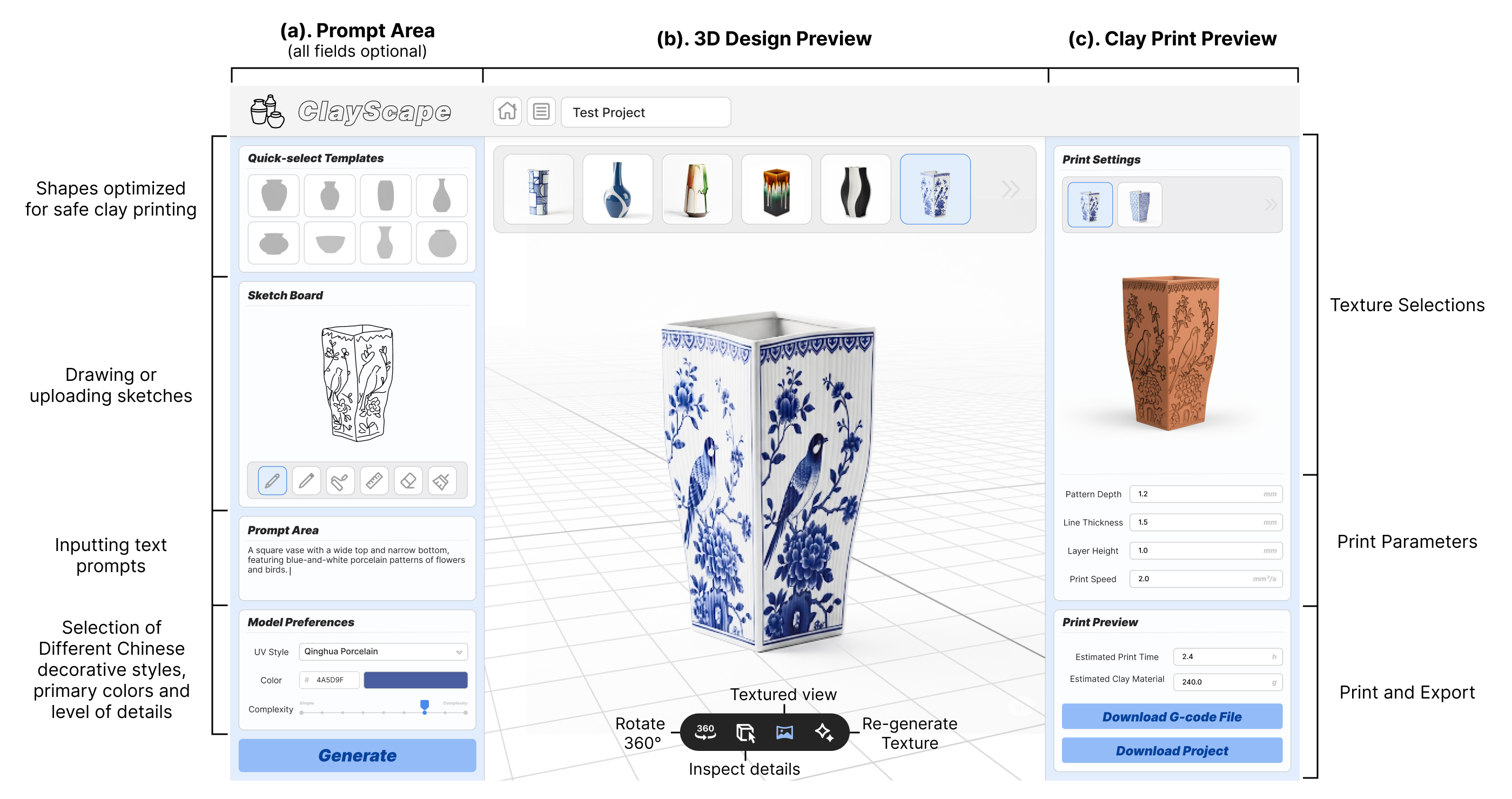}
 \caption{ClayScape Tool Interface: (a) AI prompt area for inputting different types of prompts; (b) 3D design preview for visualizing and refining the AI-generated model; (c) Clay print preview for adjusting print parameters and preparing the model for 3D printing.}
 \label{interface}
\end{figure*}

The interface of ClayScape consists of three interconnected workspaces: an AI prompt area, a 3D design preview, and a clay print preview (see Fig.~\ref{interface}). The layout and workflow design, including parameter settings, interaction flow, and tool options, is informed by insights from Phase 1.

The AI prompt area, which is in Fig.\ref{interface}.(a), serves as the primary input interface, allowing users to guide the generative AI to create ceramic designs. Based on Author A's suggestion to combine various input methods for better results, this area allows for the drawing or uploading of sketches, supports text prompts, and provides a selection of quick-select templates for shapes and patterns. This combination of tools helps users, especially those inexperienced with generative AI, to more effectively articulate their creative vision. 

\begin{figure*}[htbp]
 \centering
 \includegraphics[width=1\textwidth]{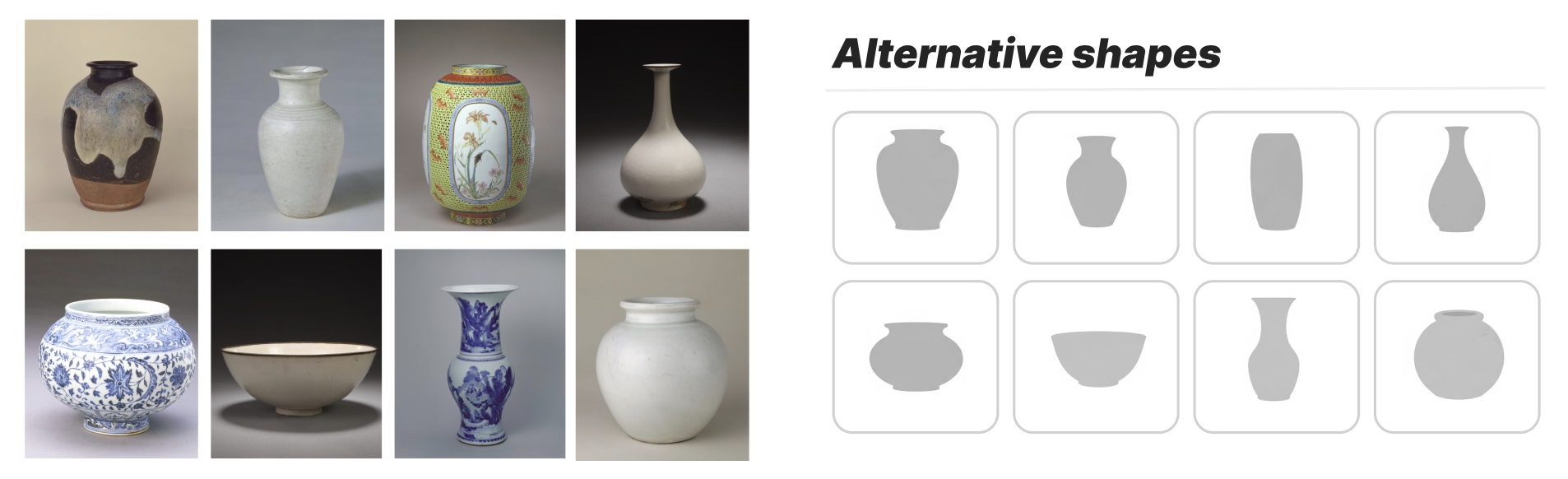}
 \caption{Alternative ceramic shape references and derived templates.
Left: Representative ceramic vessels sourced from the Palace Museum’s official collection website, including (top row, left to right) Hua porcelain jar, white-glazed bottle, floral-pattern lantern-shaped holder, and white-glazed jar; and (bottom row, left to right) blue-and-white patterned jar, white-glazed bowl, blue-and-white phoenix-tail vase, and Yuhuchun vase.
Right: Simplified and structurally adjusted shape templates derived from these historical forms, adapted to accommodate the material and mechanical constraints of clay 3D printing. }
 \label{shape}
\end{figure*}

\begin{figure*}[htbp]
 \centering
 \includegraphics[width=1\textwidth]{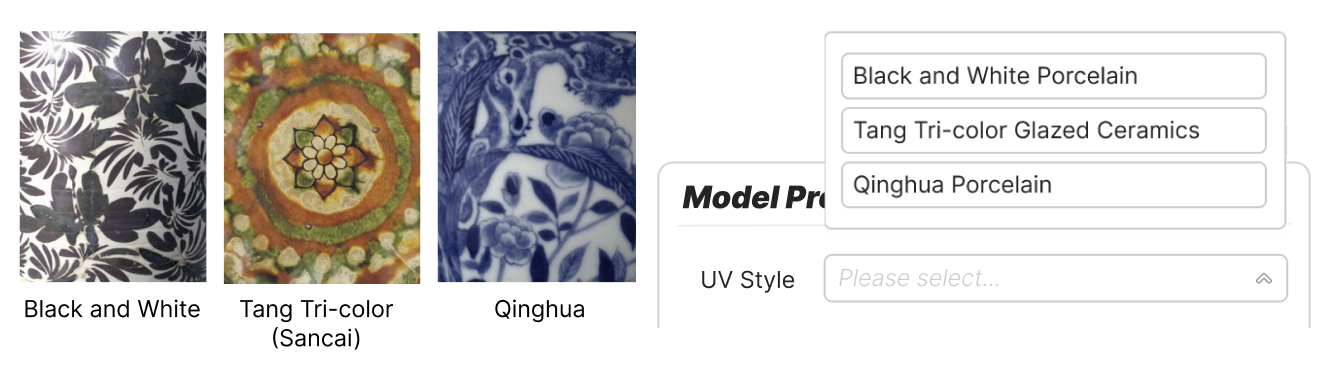}
 \caption{Left: Representative decorative motifs drawn from historically established Chinese ceramic styles, including black-and-white porcelain, Tang tri-color (Sancai) glazed ceramics, and Qinghua (blue-and-white) porcelain, used as visual references during Phase 1 introduced by Author B.
Right: Corresponding selectable pattern styles embedded in the system interface, allowing users to apply culturally grounded surface motifs to generated 3D models. }
 \label{pattern}
\end{figure*}

Furthermore, to address Author A and B's concerns regarding the physical limitations of clay printing, the system provides a set of vessel templates derived from classic Chinese ceramic forms, such as jars, vases, bowls, and bottles (Fig.\ref{shape}). These templates were selected as culturally grounded references familiar to ceramic practitioners. Rather than functioning as fixed molds, these shapes act as optional structural guides that help users anticipate printability and reduce the risk of collapse or deformation during printing and firing. Users are not necessarily required to adopt these templates, but their availability offers an accessible starting point for users with limited experience in digital modeling or clay printing. Similarly, the pattern templates provided in the system are drawn from historically established Chinese ceramic motifs (Fig.\ref{pattern}). These patterns align with the visual references introduced by Author B in 4.2.1. By embedding these motifs as selectable pattern options, the system supports culturally situated design exploration of Chinese ceramic.

The 3D design preview in Fig.\ref{interface}.(b) is the central workspace where the user can inspect the AI-generated model. This dynamic preview allows creators to rotate the design 360° and inspect details from all angles, providing immediate visual feedback on the design. The ability to re-generate texture also allows for rapid iteration based on the initial prompt inputs, helping users refine the aesthetic details of their work.

The clay print preview shown in Fig.\ref{interface}.(c) focuses on the final preparation of the model for physical creation. This section incorporates insights from Author B's experience with converting sketches to engraved models. It includes texture selections and a suite of print parameters such as Pattern Depth, Line Thickness, and Print Speed, directly referencing Author B's insights that an engraving depth of 1–1.5 mm is optimal for guiding glaze and that patterns with clear outlines work best for surface decoration. The workspace also displays a print preview with an estimated print time and clay material usage, enabling users to analyze the structural feasibility of their designs before proceeding to the final steps of print and export, which generates the necessary G-code file for clay printing.

\section{Phase 2: Creations by Crafts-people}\label{sec:p2 Findings}
During Phase 2, four participating creators engaged with the workflow to produce their own ceramic pieces. Their approaches reflected different levels of experience, ranging from beginners seeking new forms and patterns to experienced artists exploring functional transformation and storytelling. The following subsections summarize each creator’s design intention, co-creation process with ClayScape, and final outcomes.

\subsection{A01: Ice Crackle Glaze Bowl}
\begin{figure*}[htbp]
 \centering
 \includegraphics[width=0.8\textwidth]{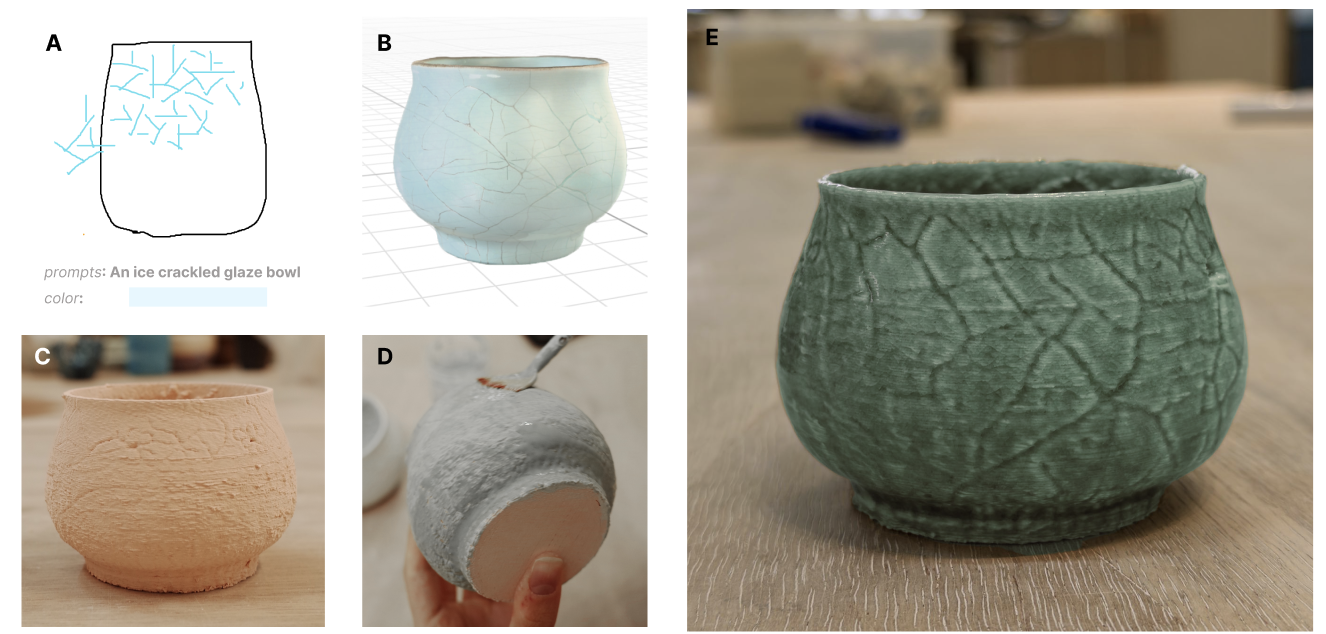}
 \caption{A01’s working process and final creation: (A) initial sketch with prompts; (B) design generated with ClayScape; (C) 3D clay-printed piece; (D) glazing in progress, and (E) final fired piece.}
 \label{A01}
\end{figure*}

A01, who had previously trained in entry-level ceramic course, wanted to attempt something beyond her beginner level, like forms she could not finish by hand and patterns she could not draw herself. She was fascinated by ice crackle glazes, and inspired by the Phase 1 results to explore whether engraving patterns could imitate crackle effects. In the design session, she created a bowl with engraved crackle textures in ClayScape. The final clay 3D printed bowl showed natural and visible crackle lines. While the final glaze did not produce a true crackle, the engraved textures themselves worked convincingly (see Fig.\ref{A01} (E)).

\subsection{A02: Floral Vase}
\begin{figure*}[htbp]
 \centering
 \includegraphics[width=0.8\textwidth]{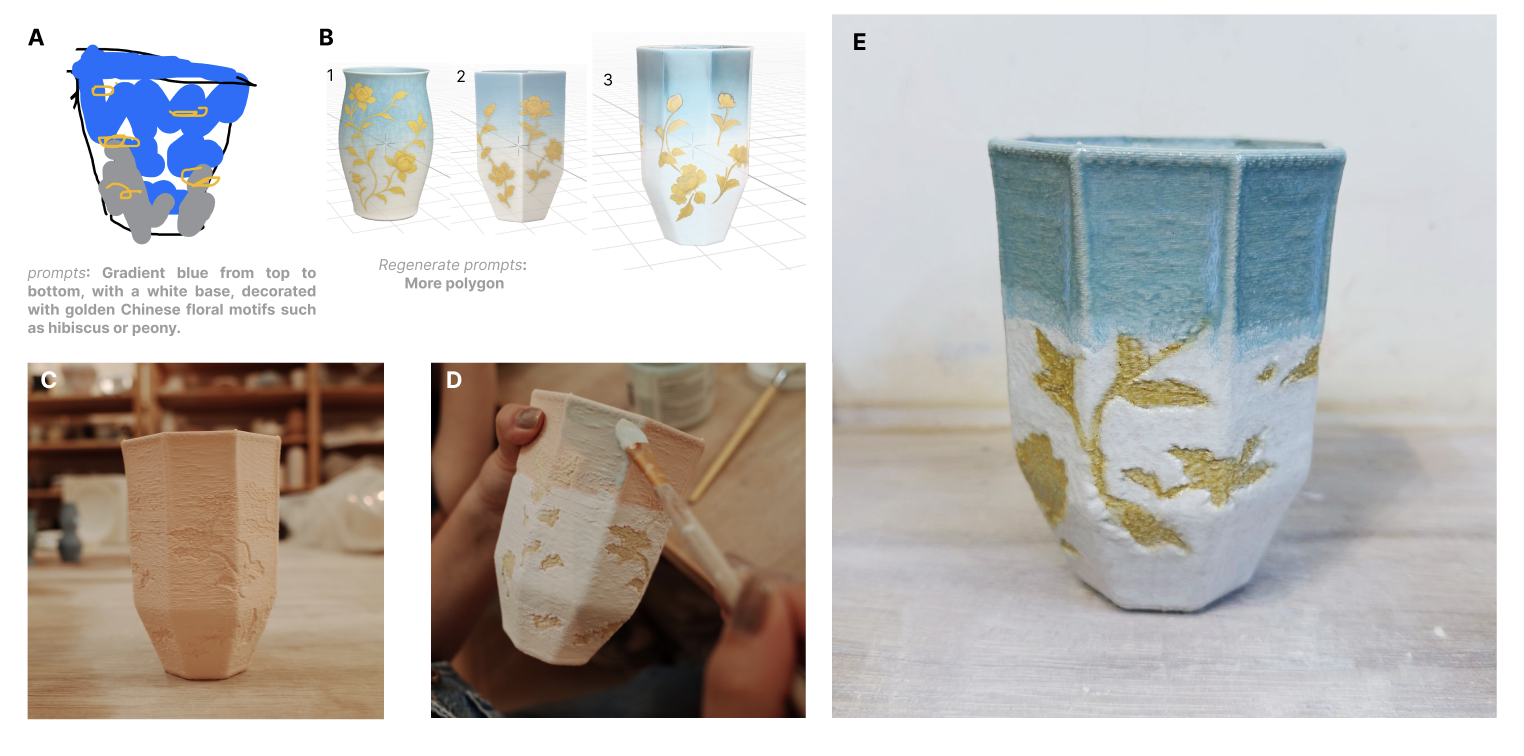}
 \caption{A02’s working process and final creation: (A) initial sketch and text prompts; (B) ClayScape-generated vase designs with different polygonal variations : 1) rounded, 2) quadrilateral, and 3) octagonal, with A02 selecting the octagonal form for its resonance with Chinese aesthetics; (C) 3D clay-printed piece; (D) glazing process guided by engraved floral textures; and (E) final fired piece.}
 \label{A02}
\end{figure*}

Also a beginner, A02 wished to create a more complex form than she had achieved before. She had some experience with gradient glazing but lacked confidence in drawing floral motifs. During the design session, she regenerated several versions of a vase with different polygonal shapes, from round to square to octagonal (Fig.\ref{A02}.(B)). She finally selected the octagonal form, which she felt resonated more with Chinese aesthetics and represented a shape she could not achieve by hand (Fig.\ref{A02}. (B)-3). Building on this form, she designed a blue-and-white gradient vase with golden Chinese floral patterns. In the glazing session, she followed the engraved texture as a guide and successfully produced a piece closely resembling the AI-generated design.

\subsection{A03: Candle Warmer}
\begin{figure*}[htbp]
 \centering
 \includegraphics[width=0.8\textwidth]{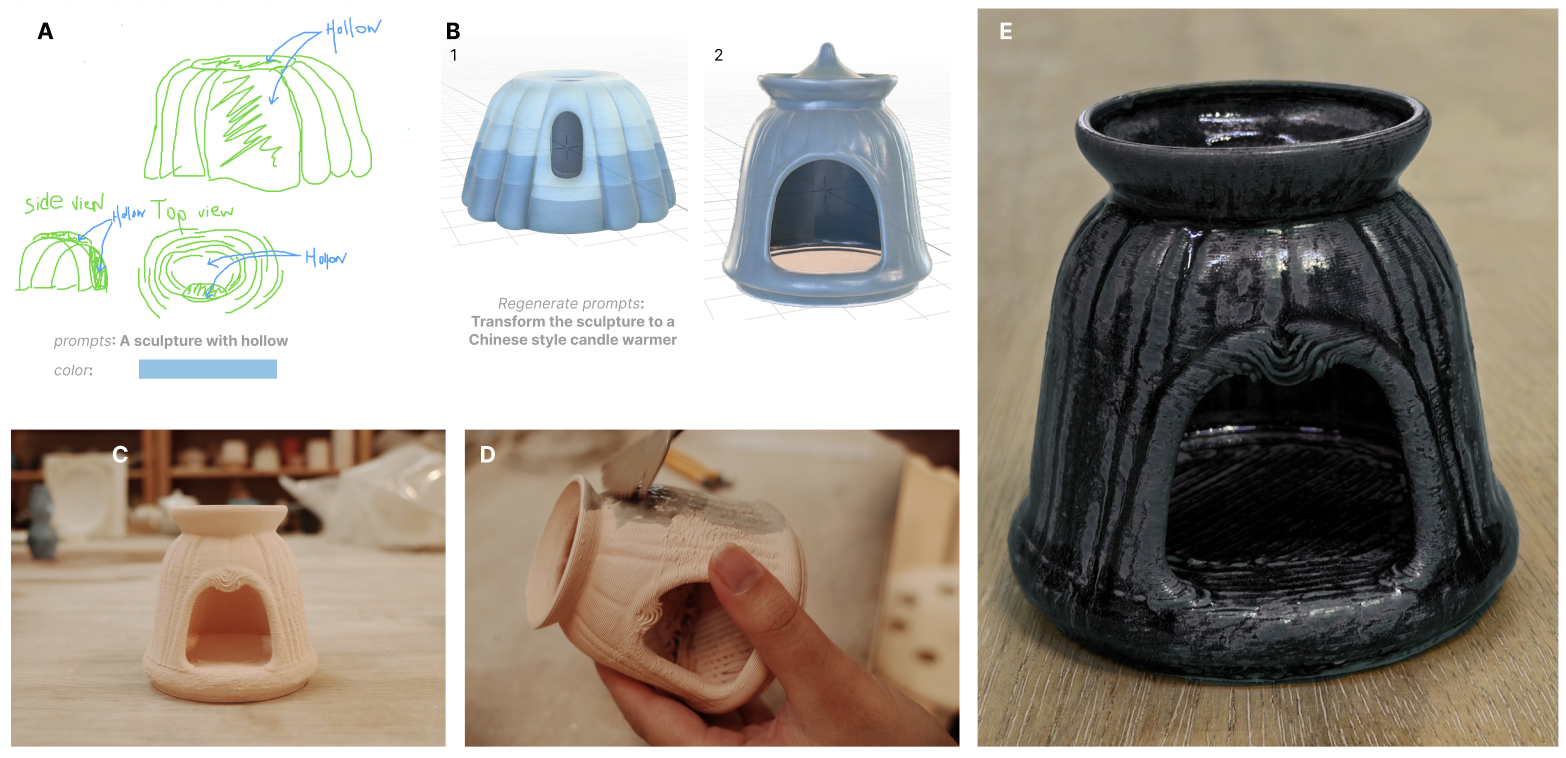}
 \caption{A03’s working process and final creation: (A) initial sketch and prompts for a hollow sculpture; (B) ClayScape-generated designs showing the transformation into a Chinese-style candle warmer; (C) 3D clay-printed piece (D) glazing process; and (E) final fired piece.}
 \label{A03}
\end{figure*}

As an experienced ceramic artist, A03 first provided a detailed sketch of a artistic sculpture. During the design session with ClayScape, she was inspired by generated results, and then regenerated to transform the sketch into a functional piece, a Chinese-style candle warmer for fragrance. The top section posed both functional and structural risks so it was removed. The final one was both functionally and visually like a Chinese traditional candle warmer (Fig.\ref{A03}.E).

\subsection{A04: Vase As A Money Jar}
\begin{figure*}[htbp]
 \centering
 \includegraphics[width=0.8\textwidth]{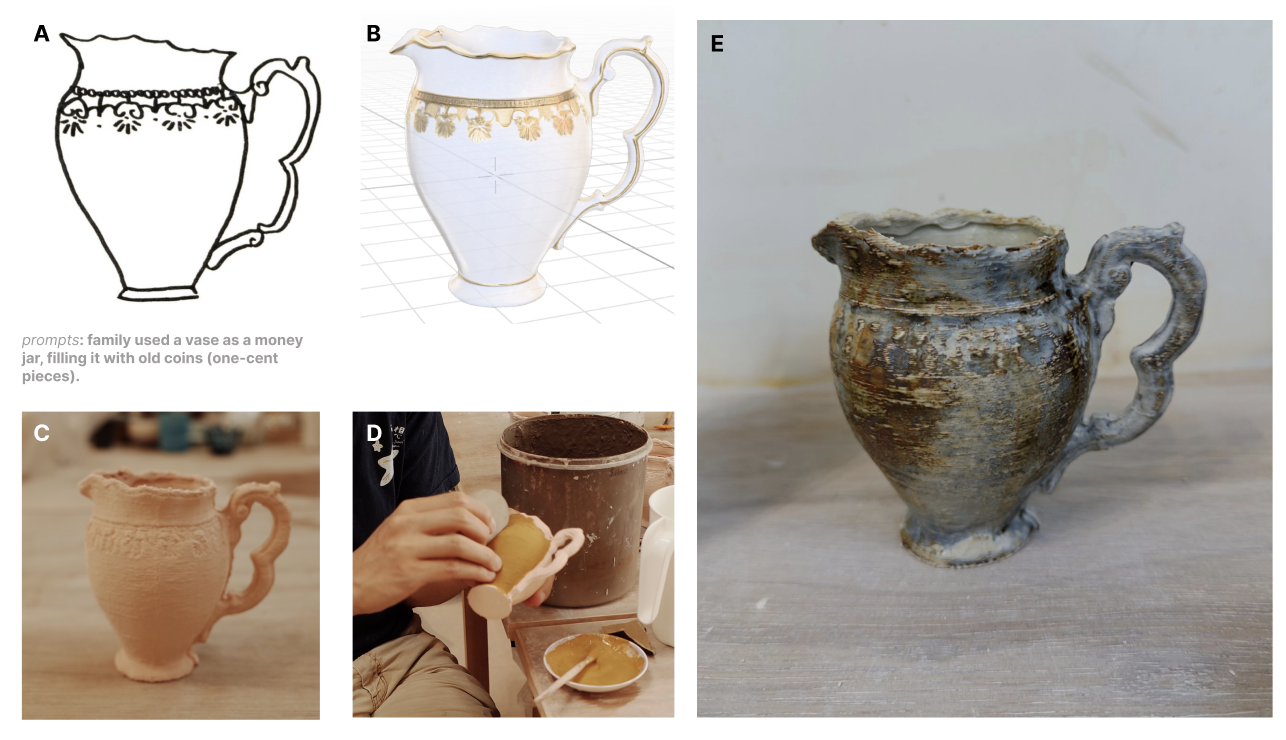}
 \caption{A04’s working process and final creation: (A) initial sketch and text prompts inspired by a family memory of using a vase as a money jar; (B) ClayScape-generated design; (C) 3D clay-printed piece; (D) glazing process using handcraft techniques to adjust decoration; and (E) final fired piece resembling an aged ceramic object.}
 \label{A04}
\end{figure*}

A04 had previously collected simple sketches and stories of old ceramic objects and transformed them into 3D models using CAD and CAM techniques. In this study, he wanted to test how ClayScape could support a similar workflow. During the design session, he redrew sketches and entered stories as text prompts to generate images and models. The chosen piece, a vase once used as a family coin container, carried the story of its contributor. In the glazing session, A04 sought to align the surface decoration more closely with his own imagination rather than relying only on the AI-generated design. The final result resembled an aged ceramic object (Fig.\ref{A04}.(E)).

\section{Phase 2: Findings on Creator Engagement and Experiences}\label{sec:p2 Findings}
\subsection{Enabling creative possibilities in ceramics making}
Feedback from the creative process and follow-up interviews indicated that creators perceived the workflow as providing tangible support compared with the more complex steps of traditional handcraft methods. Both beginners and experts recognized these benefits, though from different perspectives.

\subsubsection{Improving time efficiency}
The most obvious improvement concerned the time required to complete a piece. For beginners, the workflow reduced the effort spent planning and executing forms and surface decoration. Normally they would need to first decide on a form, sketch patterns by hand, and then attempt to transfer them onto the surface, a process that consumed significant time and sometimes led to wasted effort due to the lack of skills. Using our workflow, creators could describe their ideas for patterns to ClayScape and receive a printed piece with engraved guides, eliminating the need for preliminary hand drawing. As A02 explained: “\textit{The time efficiency is much higher. Now, in less than two hours, we already finished all the glazing.}”

For ceramic experts, the digital fabrication components were perceived more as tools or assistants, saving time and labor even if not directly enhancing creativity. A04, who had prior experience using CAD and CAM to build 3D models from sketches, compared this practice experience with his usual workflow. He noted that the ability to move from a hand sketch to an STL file meant bypassing many intermediate modeling steps: “\textit{If I have the image and the STL 3D model file, it saves a lot of time and labor. I could also have time to add some other things to find outcome.}” For him, this efficiency did not necessarily change the creative direction of his work, but it streamlined the technical preparation in some projects.

\subsubsection{Encouraging New Directions, Confidence, and Approaches}
Another contribution of the workflow was its ability to open new creative possibilities. The design sessions with ClayScape were not simply about visualizing ideas; for many participants, especially those without a clear starting concept, the randomness and unpredictability of GenAI became a source of inspiration. For example, A03 was inspired by the first generated version of her sketch, the “pudding-shaped” sculpture (see Fig.\ref{A03}. (B) - 1). It suggested the idea of transforming the shape into a functional candle warmer, a direction she had not considered before.

The workflow also expanded the scope of what creators could attempt. A01, who had prior experience using Grasshopper to build 3D models, noted that most beginners are usually restricted to parametric forms. In contrast, GenAI opened a wider pathway that allowed her to sketch irregular or unconventional shapes, which ClayScape could transform into viable ceramic designs. She explained, “\textit{My hand-drawn lines often look shaky, making the vessel seem irregular. But after the AI processed it, that randomness and irregularity were preserved as a solid effect. I used to believe that if I drew a cup, the AI would simply generate its own standard interpretation of a cup, which would make me feel boring. But it actually kept the random qualities and turned them into something lively, that was really fun.}”

A04 also got inspiration for teaching from the process. In his ceramic courses, introducing 3D clay printing usually requires teaching students modeling techniques first, which can be a barrier for those with a handcraft background and little CAD/CAM experience. After participating in this practice process, A04 saw the potential of using GenAI tools as an entry point, enabling students to design and model quickly so they could focus more on experiencing clay printing itself.

At the same time, the AI-generated designs conveyed a sense of authenticity through their resonance with Chinese traditional ceramic aesthetics, which encouraged artists’ working confidence. For A01, the engraved ice-crackle textures were inspired by classic crackle glazes, a long-established style in Chinese ceramics. Without the technical knowledge of firing and glazing required to achieve this effect, she had never considered making one before, but the mimic textures gave her confidence to attempt it. A02’s vase incorporated a gradient blue-and-white base combined with golden floral motifs, which she perceived as closely aligned with traditional decorative patterns yet made possible for her through GenAI support.

Both A01 and A02 also emphasized how the generated motifs lowered the barrier to trying complex designs. A01 said: “\textit{I cannot paint sophisticated surfaces. I learned a little Chinese traditional ink painting in primary school, but I couldn’t achieve something like patterns on blue-and-white porcelain.}” Similarly, A02 reflected: “\textit{With traditional methods, I would never have designed such complex patterns, and I was even afraid of ruining a carefully formed vessel. But this time I dared to try a more complicated piece, because I didn’t have to hand-build or model everything myself. AI and 3D printing already produced the basic vessel shape. That was really encouraging.}” These works show how the hybrid workflow could generate pieces that connect with cultural traditions while still reflecting individual imagination.

Besides inspiring beginners, it also encouraged expert A04 to experiment with innovative hand-glazing approaches on the printed surfaces. He explained: “\textit{I’m drawn to the texture, and then I try to highlight this with glazing or decoration techniques. Because the 3D printing already created some texture based on the decoration, I wanted to see how I could use this uneven surface to enhance the effect.}” His reflections also suggested a further possibility: applying traditional craft techniques back onto digitally fabricated forms, blending manual skill with AI-generated structures.

\subsection{Challenges encountered during practice}
Although the workflow enabled new creative possibilities, it also introduced challenges for participating creators. 

\subsubsection{Material and Process Constraints}
The printed surface introduced challenges for glazing. Thin printed clay layers created small gaps, and engraved textures left more roughness. Beginners in particular found it difficult to glaze evenly. A01 reflected: “\textit{The printed stacking on the surface wasn’t very even, maybe because of the patterns. Those parts made it hard to apply the glaze uniformly.}” Similarly, A02 compared the printed outcomes to her handmade work: “\textit{Although the form looks better than what I made by hand, the printed surface is hard to glaze if not polished. I had to apply many coats to avoid uneven results.}”

\subsubsection{Feasibility Challenges in Clay Print}
\begin{figure*}[htbp]
 \centering
 \includegraphics[width=0.8\textwidth]{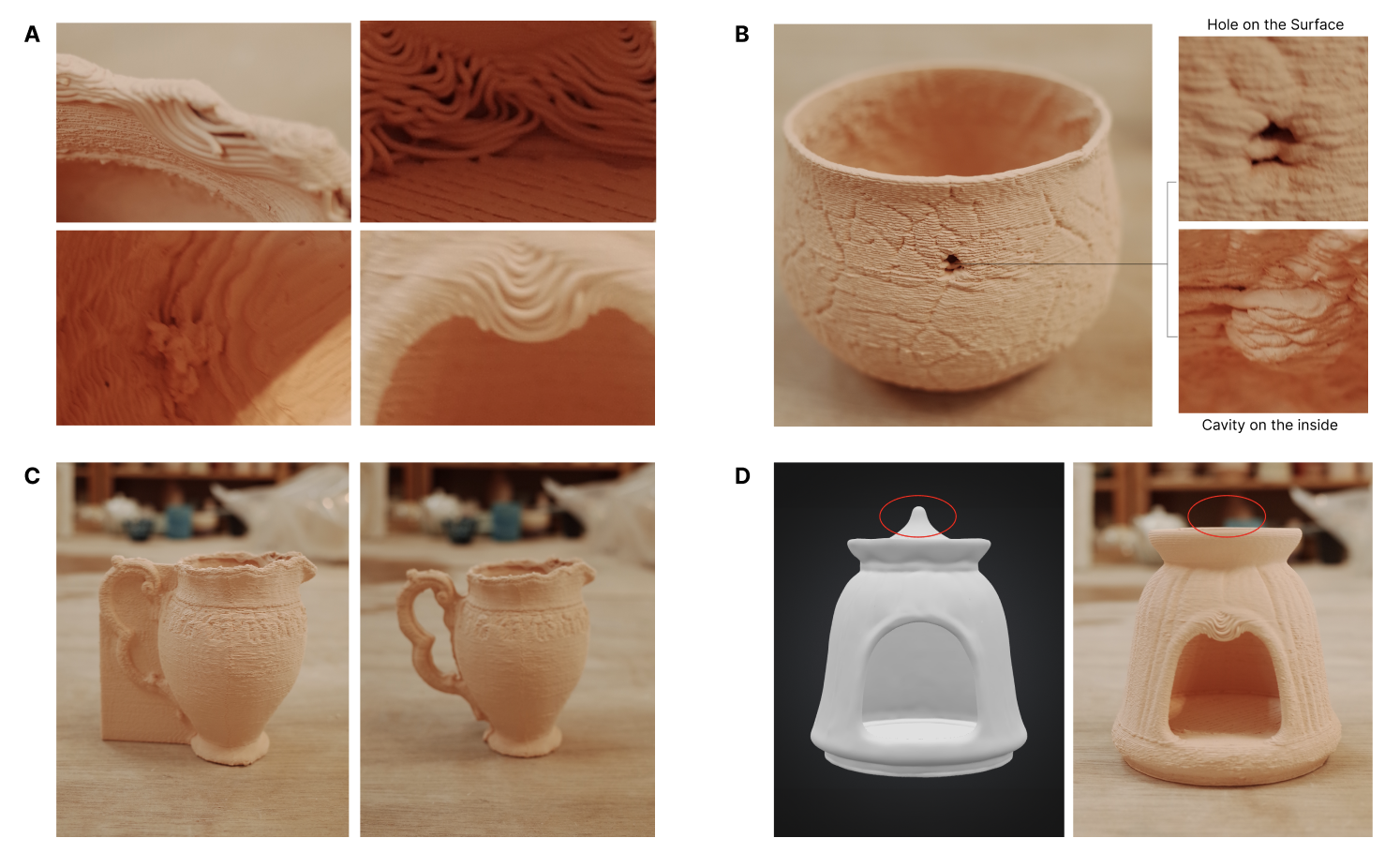}
 \caption{Structural feasibility issues observed in clay printing: (A) Examples of collapsed or irregular layering in printed clay; (B)A01’s bowl showing an unexpected hole on the surface and a corresponding cavity on the inside; (C) A04’s vessel requiring added support for the handle during printing; (D) A03’s candle warmer, with the sharp lid (highlighted by red circle) removed to reduce risk of failure.}
 \label{feasibilities}
\end{figure*}

Despite the improvements informed by Phase 1, including the addition of suggested template shapes in the input workspace and print preview settings to reduce print failures, ClayScape still sometimes produced models at risk during clay printing. There are some minor collapsing parts on printed final pieces (see Fig.\ref{feasibilities}.(A)). AI-generated models also have not controllable effects may affect printing. The bowl of A01 had an unexpected hole in the wall (see Fig.\ref{feasibilities}.(B)). Even though she usually embraced randomness, she found this unacceptable: “\textit{I really like the randomness of creation, AI generation and clay print both have it. For example, when printing a head, the collapsed clay might look like hair, which is reasonable. But this hole was not reasonable.}”

Beginners tended to follow template shapes and thus avoided printing risks, but experts intentionally pursued more complex designs that frequently challenged feasibility of printing. In particular, A04 acknowledged that some forms are not printable or require parts to be produced separately, but he still wanted to try the challenging designs rather than the simple boring ones. This situation often required Author B’s intervention to ensure feasibility, such as removing the sharp lid of A03’s candle warmer and adding support to the handle of A04’s vessel (see Fig.\ref{feasibilities}.(C) and (D)) before printing. However, this technical support made A04 reflect: “\textit{If creators still need to add something or divide some parts for printing complex forms, then we go back to the earlier question, creators must be equipped with CAM skills and knowledge of structural feasibility in clay print. So it is still limited.}” 

\subsubsection{Compromising Creative Freedom}
Participating creators felt their creative freedom was restricted at several points in the workflow. A03 admitted she was unsure how to adapt her usual crafting practice to the AI’s capabilities, while she's not familiar with it. She pointed out that the AI-generated model of her candle warmer included only a very small hollow (see Fig.\ref{A03}.(B)-1), raising doubts about whether a candle would fit. This level of detail, which she could easily adjust by hand, was not yet controllable through ClayScape. For her, if simple designs could be easily hand-made and complex ones risked failure, then this hybrid workflow added limited value.

Participants noted that the workflow’s process of translating ideas into clearly visualized and fabricated forms could reduce the sense of agency in their creations. A01 commented: “\textit{It was without the unpredictability of handcraft.}” Similarly, A02 remarked: “\textit{It looks more like something for mass production. It could be a craft piece, but not as art.}" She further explained that making sketch-like designs overly visualize reduced their expressiveness: “\textit{Some ideas about ceramics are rough and spontaneous. Making them too concrete takes away their wild beauty.}”

In addition, while engraved patterns initially helpful as glazing guides, sometimes constrained the creative decorations. A01 noted: “The engraved UV patterns tell me where to apply color, but sometimes during glazing I had new ideas.” A04 questioned the artistic value of tracing generated motifs: “\textit{It’s kind of taking away the quality of the painting. You’re tracing a pattern, but it’s still not a painting. It’s a generated painting, a secondary painting.}” 

Although these constraints revealed clear tensions in practice, they did not diminish the creators’ motivation, especially the beginners. They even suggested improvements for regaining creative freedom: A01 proposed adding varied depths to the engraved patterns to allow greater flexibility, while A02 suggested using removable stickers or templates as glazing guides, which could support decoration without locking it into fixed engravings. Those feedback reflected the engagement and expectation in this hybrid fabrication workflow, which we discuss in the next subsection.

\subsection{From Engagement to Expectation: Creator Experience in Hybrid Practice}
While A03 reported more confusion than engagement when functional control was not sufficiently exposed, the other participants described the workflow as both sustaining their engagement and influencing their expectations. For beginners in particular, engagement was strong and motivating, grounded in the sense of achievement when their designs became tangible. At the same time, the workflow also created anticipation: participants expressed imaginations for future possibilities.

\subsubsection{Engagement Through Achievement}
For beginners, the workflow fostered strong feelings of engagement, largely because it enabled them to achieve results they had previously thought beyond reach. The reduction in complexity compared with traditional crafting did not lessen their sense of accomplishment but amplified it instead. A01 explained: “\textit{When AI presented my design, I felt a sense of achievement, even a sense of belonging.}” She emphasized that her engagement with AI-assisted work was no less than with handcrafting.

A02 expressed similar feelings, describing how the workflow gave her confidence to attempt outcomes she had once considered possible only for masters: “\textit{Before, I thought only masters could make such works, but now I feel I also have hope to complete them.}” These experiences suggest that engagement improve not only from the complex steps but also from the sense of empowerment that the hybrid fabrication provided.

\subsubsection{Expectation and Ownership in the Process}
Beyond immediate achievement, the workflow also shaped participants’ expectations and sense of ownership. A01 noted that only receiving an AI-generated preview would not have felt meaningful to her. Instead, her engagement deepened when she realized her design could be physically realized: “\textit{If AI only generated an image for me, I would not feel it was mine. But once I knew it could be made into a real piece, I started to feel anticipation.}” Reflecting on this shift, she added: “\textit{In traditional craft, I felt I was shaping a work from 0 to 1. Now it feels like influencing from 1 to 2.}”

A02 also connected engagement with future anticipation, imagining how she could extend the workflow to new directions: “\textit{I can already imagine using this workflow to make small sculptures of cats or dogs. AI and 3D printing might even restore them more easily than hand-building.}”

For A04, expectation extended beyond his own practice to broader participation. Besides the teaching considerations mentioned in 7.1.2, he saw the workflow as a way to improve previous participatory ceramic projects and engage more non-professional people in without being limited by the time and labor of producing large numbers of models himself: “\textit{If I want to engage more people, this workflow can help me improve my previous project. I don’t need to worry about building too many models by myself.}”

These reflections highlight how the workflow encouraged engagement and revealed its potential to extend into participants’ future practice. At the same time, they inspire researchers to refine the workflow itself, seeking ways to balance traditional handcraft values with new digital practices, and to optimize the capabilities of AI tools so they better support both creative freedom and material feasibility.

\section{Discussion}\label{sec:Discussion}
\subsection{Interplay of GenAI and Digital Fabrication in Craft}
Although prior projects have combined CAM-based design methods with clay printing \cite{bourgault_coilcam_2023}, or applied GenAI in other areas of digital fabrication \cite{tao_aifiligree_2025}, they have not explored this particular hybrid workflow. Our findings from two studies show that GenAI and clay printing can mutually support each other in craft practice, creating a productive interplay.

\subsubsection{GenAI Supporting Digital Fabrication}
Previous studies demonstrated that the visualization capabilities of AI tools enhance the efficiency of the design process \cite{tao_aifiligree_2025}. We observed similar effects: participants recognized that AI-aided tools accelerated visualization and 3D modeling, improving efficiency in both design and fabrication stages. However, the impact extended beyond efficiency.  Similar to CAM-based systems in previous research projects \cite{frost_sketchpath_2024}, GenAI lowered entry barriers for handmaking-based creators to engage with clay printing, which normally requires prior computational design knowledge. For example, A02, who was unfamiliar with clay 3D printing, described how the workflow made the technology feel tangible and usable rather than distant or inaccessible. In this sense, GenAI opened an avenue for handcraft creators to access digital fabrication.

More importantly, GenAI also helped blur the boundaries between digital fabrication and traditional crafts. This aligns with prior HCI arguments that computational methods can connect digital and traditional practices \cite{Jacobs_computational_2025}. Author B had previously expected that clay printing could move beyond geometric outputs to produce forms resonant with traditional ceramic aesthetics. ClayScape was trained on traditional craft visual references, so it could generate design with crafted looks. As we argued in Section 8.1, digital fabrication should not be seen as a replacement for traditional craft but as an assistive tool. We extended this trajectory by embedding GenAI into the workflow, making traditional aesthetics more accessible within digital practice.

\subsubsection{Digital Fabrication Supporting AI-aided Design}
Most applications of GenAI in craft remain at the design stage, focusing on computational generation without testing whether outputs can be physically realized \cite{wang_harmonycut_2025, tao_aifiligree_2025}. There is an uncertainty that may not every design be feasible by hand craft skills, so how AI-generated outputs might translate into physical form left as a problem. By combining clay printing with GenAI, our workflow advanced AI-aided design into a tangible stage. While digital fabrication can act as a bridge between computational design and craft-making \cite{Jacobs_digital_2016, Hirsch_Nothing_2023}, it also enables participants to materialize and evaluate their AI-generated ideas.

This transition from digital to physical was important for sustaining engagement. The relationship between creators and their work is not fixed but continually reconfigured through interaction with tools and materials \cite{suchman1993working}. If AI-generated designs cannot be made tangible, anticipation would quickly diminish and even cause aesthestic fatigue \cite{AI_Jiang_2023}. As A01 said that the successful physical realization of her AI-generated design increased both her expectations and her interest in AI tools. The finding demonstrates that clay printing as not only a partner to GenAI but also an essential step in pushing AI-aided design to the next stage.

\subsection{Hybrid Fabrication in Ceramic Craft Practice}
Hybrid digital fabrication workflows in traditional crafts are increasingly explored in HCI and design fields, aiming to improve efficiency and open new possibilities for craft practice by combining computational design with digital production \cite{Jacobs_digital_2016, Jacobs_computational_2025, zoran_Hybrid_2015, Hirsch_Nothing_2023}. Several domain-specific CAM systems for clay printing have been developed \cite{bourgault_coilcam_2023, frost_sketchpath_2024, friedman-gerlicz_weaveslicer_2024}. Our workflow builds on this line of work by integrating GenAI at the design stage, using ClayScape as the CAM-based tool, and applying clay 3D printing at the fabrication stage within a hybrid process.

\subsubsection{Non-Parametric Approaches for CAM Design}
Earlier clay printing methods and 3D printing hybrid design relied on parametric tools \cite{PotScript}, but recent work such as SketchPath demonstrated how drawing-based input could lower the barriers of CAM expressions through non-parametric design \cite{frost_sketchpath_2024}. That project also suggested that non-parametric approaches may better align digital fabrication with manual craft practices. 

We followed this direction when developing ClayScape by enabling creators to sketch ideas by hand in addition to using text or parameters. Our findings similarly showed that non-parametric input allowed participants to design more freely. The combination of hand sketch and text descriptions preserved their sense of authorship and engagement rather than reducing creativity to parameter tuning (7.3.1, 7.3.2). As A01 reflected, the tool expanded shape and texture possibilities beyond those of parametric design (7.1, 7.2). 

In addition, the visualization and rendering capabilities of GenAI provided a more comprehensive design preview, despite limited controllability on current AI component, it enables creators to test and refine ideas more easily and efficiently. Thus, while we build on prior non-parametric workflows, our contribution lies in extending them with GenAI-driven visualization, which strengthen design feedback and creative confidence.

\subsubsection{Reflections on Action-Oriented Fabrication}
Bourgault et al. define action-oriented fabrication workflows as processes in which craftspeople create artifacts through repeated actions on materials \cite{Developing_Action-Oriented}. They adopt this workflow in digital fabrication to build new systems, aligning parts of the CAM process with human handcrafting steps to foster collaboration between machines and craft \cite{bourgault_coilcam_2023}. This is something we did not incorporate extensively in our hybrid fabrication workflow. We did not consider the similarity with hand craft movements, as our initial intention was to reduce the complexity of working steps for creators. As discussed in Section 7,  entry-level participants such as A01 and A02 were more interested in creating designs they could not easily complete by hand. We also worried that making the process too similar to hand-building might discourage beginners.

However, since we already employed GenAI and clay printing in the fabrication process, we were cautious that reducing too many elements of manual craft could diminish creators’ sense of engagement. In our approach, we retained several mandatory procedures from traditional craft practice, such as hand sketching and glazing. While the hybrid fabrication workflow may distance creators from direct interaction with materials, we expected that retaining hands-on steps such as glazing and decorating would enable them to remain physically and experientially connected to their work. During the process, we found that although material constraints in the glazing process introduced challenges (Section 7.2.1), they also enhanced creators’ sense of engagement in making their work. This hybrid workflow indeed helps preserve participants’ sense of ownership while increasing their anticipation compared to purely GenAI-driven design (7.3.2).

At the beginning of this project, we expected ClayScape and clay printing could serve as replacements for some traditional steps. However, our findings suggested that they are better positioned as assistive tools rather than substitutes, in order to balance digital technologies and craft practices. The primary goal of creators is to realize an envisioned piece. While the hybrid workflow enables the translation of ideas into tangible forms, participants also noted that this process reduce the unpredictable beauty associated with manual craft (7.2.3).  The use of GenAI-supported design and clay printing does not imply that creators must fully shift to hybrid digital fabrication workflows.  A04, who had expertise in both hand-building and clay printing, emphasized that the two should not replace each other, as each has distinct value. This reflection, informed by action-oriented fabrication, makes us reconsider the relationship and balance between human and machine in hybrid workflows.

\subsubsection{Embedding Cultural Parameters for Hybrid Fabrication}
Our study also pay attention to the role of cultural specificity in shaping hybrid workflows. Chinese ceramic practices illustrate how culture, vessel forms, surface ornamentation, and material processes are deeply intertwined \cite{valenstein1989handbook}. This context highlights the importance of interpreting and translating culturally meaningful patterns and forms into our workflows, which is an aspect that has recently begun to receive more attention in digital fabrication \cite{wang_harmonycut_2025, tao_aifiligree_2025}.

Silva Lovato et al. showed that cultural contexts shape ceramic practice as much as technical systems \cite{silva_lovato_american_2025}. Building on this insight, our workflow differs from previous CAM-based clay printing systems by embedding cultural parameters such as Chinese glaze palettes, traditional motifs, and archetypal shapes into the design process (Section 5). These parameters, however, are optional rather than prescriptive. They serve as references that creators can draw on while retaining the freedom of non-parametric design. None of the participants felt strongly constrained by these options, as they are not mandatory; instead, participants found them useful at times for generating more accurate and culturally resonant designs. This demonstrates that hybrid fabrication can support ceramic design by providing not only technical support but also cultural grounding and creative authenticity.

\subsection{Creative Collaboration in Hybrid Fabrication}
Artist–researcher collaborations are no longer rare in HCI, as researchers increasingly recognize that artists can serve as technical collaborators \cite{toka_practice-driven_2024, Devendorf_Craftspeople_2020,lc_contradiction_2023,lc_presentation_2022,liu_dreamscaping_2024}. In digital fabrication projects related to clay printing, many studies conducted as artist residency program, creating mutual benefits for both artistic and technical practices \cite{Devendorf_Towards_2023}. Our research learned from those approaches to let the collaborations directly informed the development of the hybrid workflow.

In Phase 1, knowledge exchange between Author A and Author B was particularly formative. After Author A learned about traditional Chinese ceramic processes and aesthetics, she was able to identify which patterns and visuals were most suitable for training GenAI. Conversely, once Author B became familiar with the capabilities of GenAI, he began to see how it might connect with clay printing to lower the barriers of traditional ceramic making. These complementary insights informed the design of ClayScape, the hybrid workflow, and the subsequent creator practices.

The involvement of additional participating artists in Phase 2 also generated expectations and suggestions that offered valuable insights for future refinement. While previous artist-researcher collaborative programs that primarily involved mature ceramic artists, our study also included beginners. It allowed us to evaluate the accessibility of the hybrid workflow while at the same time comparing their perspectives with those of experienced practitioners.

In addition, while earlier studies often ran long-term artist residency programs that produced larger bodies of work and deeper self-reflection, our shorter-term session limited timeframe kept the experience fresh for participants, which encouraged them to share candid reflections and immediate feedback on the workflow. Both of collaborative approaches demonstrate how artist–researcher partnerships can expand technical exploration while remaining grounded in craft expertise.

\subsection{Limitation and Future Work}
To identify directions for future development, we must first understand the current limitation of the workflow. Even though the technical barriers is lowered and more creative exploration among participants were fostered, we cannot neglect the constraints in the system and the CAM tools. These limitations mainly stem from current technology and the nature of human-AI collaboration. The said limitations should be further improved to create a more ideal environment for the hybrid creative process between human, AI and CAM methods.

First, a significant challenge arises from the present development of clay 3D printing technology, specifically in the Direct Ink Writing process. The physical outcomes are inevitably shaped by the mechanics of the 3D printer and the material properties of clay (7.2). Print quality is limited by several factors including visible layer lines caused by limited printing resolution, structural constraints on overhang angles due to the softness of wet clay, and occasional air pockets within the material affecting the final aesthetic and structural integrity. These technological imperfections can conflict with the high standards of surface quality inherent to traditional ceramic craftsmanship which requires flawless finishes. Thus, although ClayScape was designed based on findings from Phase 1, technical challenges during clay printing could not be entirely eliminated, occasionally requiring researcher intervention to ensure print feasibility. As such, the extent to which ClayScape supports fully independent use remains a direction for future evaluation. Furthermore, while the workflow simplifies the creative process, a foundational understanding of multiple disciplines is critical to achieve high quality final pieces. Users must possess adequate basic knowledge in effective GenAI prompting, 3D printer operation and ceramic techniques of glazing and firing. This cross-disciplinary requirement represents a learning curve that remains a considerable hurdle for first-time users or craftsmen.

The integration of GenAI represents a trade-off between speed and control. The technology has accelerated the conceptualization and modeling phases, at the same time sacrificed a degree of granular control over the form compared to traditional computational assisted modeling techniques. For users who begin with a strong and specific vision, the AI-generated models can be seen as useful inspirational references but may fail to visualize the user’s exact intent, and the output might feel misaligned with the user's initial expectation (7.2.3 A03 afraid of the hollow is too small to put the candle in).

The mentioned limitations has hinted directions for future work. An immediate and effective improvement will be advancements in hardware. The development of higher-resolution extrusion systems, quick-setting and more robust clay composition, and extruder with improved de-airing mechanisms will instantly address the issue of thick layer lines, complex structural integrity and material consistency.

Other than the hardware engineering aspect, further research can be conducted on enhancing the AI component to be more responsive and predictable, which involves fine tuning datasets of traditional Chinese ceramics to improve the generative models' cultural and stylistic fidelity. The core objective here is to increase the controllability of AI, so that a more iterative and interactive dialogue between human and AI can be achieved. This improvement is critical for users to provide feedback on initial generations to generate subsequent output more precisely. By bridging the gap between AI inspired generations and users' intended design, creators can create works that fully cater their needs.

Finally, an adaptive strategy carried out by one of the experienced participant (A04) has demonstrated how expert craft knowledge could be seamlessly integrated in newer CAM process. The experienced artisan observed the state of the printed objects and applied glazing techniques that matched or even highlighted the unique features of the printed output. This observation suggested a promising future direction, which is to examine how craft knowledge can be documented and formally encoded into the digital workflow. Future systems can be designed to suit material behavior and finishing techniques, leading to a deeper synergy between the mind of craftsmen and the inspirational capabilities of computer assisted manufacturing.

\section{Conclusion}\label{sec:Conclusion}
We contribute a hybrid workflow that integrates GenAI and clay 3D printing, enabling ceramic creators to design and produce textured Chinese ceramics without requiring advanced handcraft or CAD/CAM skills. By reducing technical and procedural complexity, this workflow makes digital fabrication more accessible while improving time efficiency. Its non-parametric design process, supported by GenAI’s visualization and rendering capabilities, further expands creative possibilities and provides creators with immediate feedback on their ideas. Our findings show that creators at different levels of ceramic expertise engaged with the hybrid workflow in distinct ways. Beginners gained confidence through accessible entry points into ceramic making, while experienced practitioners critically negotiated how their craft knowledge could be integrated into emerging CAM-based processes, highlighting both the opportunities and tensions of hybrid fabrication. It offers insights into how hybrid fabrication workflows can balance accessibility with creative agency, while supporting culturally grounded ceramic practices.



\bibliographystyle{ACM-Reference-Format}
\bibliography{references}


\end{document}